\documentclass[12pt]{spieman}  
\usepackage{amsmath,amsfonts,amssymb}
\usepackage{graphicx}
\usepackage{setspace}
\usepackage{tocloft}

\title{Soft X-ray Imager aboard Hitomi (ASTRO-H)}

\author[a, *]{Takaaki Tanaka}
\author[a]{Hiroyuki Uchida}
\author[b]{Hiroshi Nakajima}
\author[b]{Hiroshi Tsunemi}
\author[b]{Kiyoshi Hayashida}
\author[a]{Takeshi G. Tsuru}
\author[c]{Tadayasu Dotani}
\author[b]{Ryo Nagino}
\author[b]{Shota Inoue}
\author[b]{Shohei Katada}
\author[a]{Ryosaku Washino}
\author[c]{Masanobu Ozaki}
\author[c]{Hiroshi Tomida}
\author[c]{Chikara Natsukari}
\author[c]{Shutaro Ueda}
\author[c]{Masachika Iwai}
\author[d]{Koji Mori}
\author[d]{Makoto Yamauchi}
\author[d]{Isamu Hatsukade}
\author[d]{Yusuke Nishioka}
\author[d]{Eri Isoda}
\author[e]{Masayoshi Nobukawa}
\author[f]{Junko S. Hiraga}
\author[g]{Takayoshi Kohmura}
\author[h]{Hiroshi Murakami}
\author[i]{Kumiko K. Nobukawa}
\author[j]{Aya Bamba}
\author[k]{John P. Doty}
\affil[a]{Department of Physics, Kyoto University, Kitashirakawa Oiwake-cho, Sakyo, Kyoto, Kyoto 606-8502, Japan}
\affil[b]{Department of Earth and Space Science, Osaka University, 1-1 Machikaneyama, Toyonaka, Osaka 560-0043, Japan}
\affil[c]{Institute of Space and Astronautical Science, JAXA, 3-1-1 Yoshinodai, Chuo, Sagamihara, Kanagawa 252-5210, Japan}
\affil[d]{Faculty of Engineering, University of Miyazaki, 1-1 Gakuen Kibanadai Nishi, Miyazaki, 889-2192 Japan}
\affil[e]{Faculty of Education, Nara University of Education, Takabatake-cho, Nara, Nara 630-8528, Japan}
\affil[f]{Department of Physics, Kwansei Gakuin University, 2-2 Gakuen, Sanda, Hyogo 669-1337, Japan}
\affil[g]{Department of Physics, Tokyo University of Science, 2641 Yamazaki, Noda, Chiba 270-8510, Japan}
\affil[h]{Department of Information Science, Tohoku Gakuin University, 2-1-1 Tenjinzawa, Izumi, Sendai, Miyagi 981-3193, Japan}
\affil[i]{Department of Physics, Nara Women's University, Kitauoya-nishimachi, Nara, Nara 630-8506, Japan}
\affil[j]{Department of Physics, University of Tokyo, 7-3-1 Hongo, Bunkyo, Tokyo 113-0033, Japan}
\affil[k]{Noqsi Aerospace Ltd, 2822 S Nova Road, Pine, CO 80470, USA}

\cftpagenumbersoff{figure}
\cftpagenumbersoff{table} 
\begin{document} 
\maketitle

\begin{abstract}
The Soft X-ray Imager (SXI) is an imaging spectrometer using charge-coupled devices (CCDs) aboard the Hitomi X-ray observatory. The SXI sensor has four CCDs with an imaging area size of $31~{\rm mm} \times 31~{\rm mm}$ arranged in a $2 \times 2$ array. Combined with the X-ray mirror, the Soft X-ray Telescope, the SXI detects X-rays between $0.4~{\rm keV}$ and $12~{\rm keV}$ and covers a $38^{\prime} \times 38^{\prime}$ field-of-view. The CCDs are P-channel fully-depleted, back-illumination type with a depletion layer thickness of $200~\mu{\rm m}$. Low operation temperature down to $-120~^\circ{\rm C}$ as well as charge injection is employed to reduce the charge transfer inefficiency of the CCDs. The functionality and performance of the SXI are verified in on-ground tests. The energy resolution measured is $161$--$170~{\rm eV}$ in full width at half maximum for 5.9~keV X-rays. In the tests, we found that the CTI of some regions are significantly higher. A method is developed to properly treat the position-dependent CTI. Another problem we found is pinholes in the Al coating on the incident surface of the CCDs for optical light blocking. The Al thickness of the contamination blocking filter is increased in order to sufficiently block optical light.  
\end{abstract}


{\noindent \footnotesize\textbf{*} Address all correspondence to: Takaaki Tanaka, E-mail:  \linkable{ttanaka@cr.scphys.kyoto-u.ac.jp} }

\begin{spacing}{1} 

\section{Introduction}
Hitomi\cite{hitomi_jatis,hitomi}, formerly known as ASTRO-H, is the sixth Japanese X-ray astronomy satellite, which was launched on February 17, 2016 aboard 
an H-IIA rocket from JAXA's Tanegashima Space Center. 
The Soft X-ray Imager (SXI)\cite{tsunemi2010,hayashida2011,hayashida2012,tsunemi2013,hayashida2014,tanaka2015,tsunemi2016} is 
an imaging spectrometer aboard Hitomi covering the energy range between 0.4~keV and 12~keV. 
X-rays are focused by the Wolter type I mirror optics, the Soft X-ray Telescope (SXT-I)\cite{sxt}, with a focal length of 5.6~m. 
An array of X-ray charge-coupled devices (CCDs) covers a large field-of-view (FoV) of $38^{\prime} \times 38^{\prime}$. 
Almost the same energy range as the SXI is covered also by the Soft X-ray Spectrometer (SXS)\cite{sxs}, which features a superb energy resolution 
of $\simeq 5~{\rm eV}$ in full width at half maximum (FWHM) for 6~keV X-rays. 
However, the $6 \times 6$ microcalorimeter array of the SXS covers only a small FoV of  $3^{\prime} \times 3^{\prime}$ and the 
imaging capability is limited. 
The SXI, with its large FoV and small pixels, plays a complimentary role. 

We developed the SXI based on our experiences through the development of the X-ray CCD cameras for astronomical use, 
the SIS on ASCA\cite{asca}, the XIS on Suzaku\cite{XIS}, and 
the SSC on MAXI\cite{ssc,ssc2}. 
One of the major improvements of the SXI from those CCD cameras is a thicker ($200~\mu{\rm m}$) depletion layer. 
Another improvement is lower ($-120~^\circ{\rm C}$) operation temperature to mitigate radiation damage effects. 
It is known that operation temperature of $-120~^\circ{\rm C}$ or lower results in lower charge transfer inefficiency (CTI)\cite{mori2013}, 
which was actually demonstrated by Chandra ACIS\cite{grant2006} and XMM-Newtion EPIC-MOS\cite{sembay2004}. 
To further decrease CTI, we employ a charge injection technique similar to that developed for Suzaku XIS\cite{uchiyama2009}. 

In this paper, we describe the design and data processing of the SXI. 
Data taken during on-ground tests are shown to present its pre-launch performance. 
We report on in-orbit performance of the SXI in a separate paper.\cite{sxi_inorbit}

\section{SXI System}
\begin{figure}
\begin{center}
\begin{tabular}{c}
\includegraphics[width=15cm]{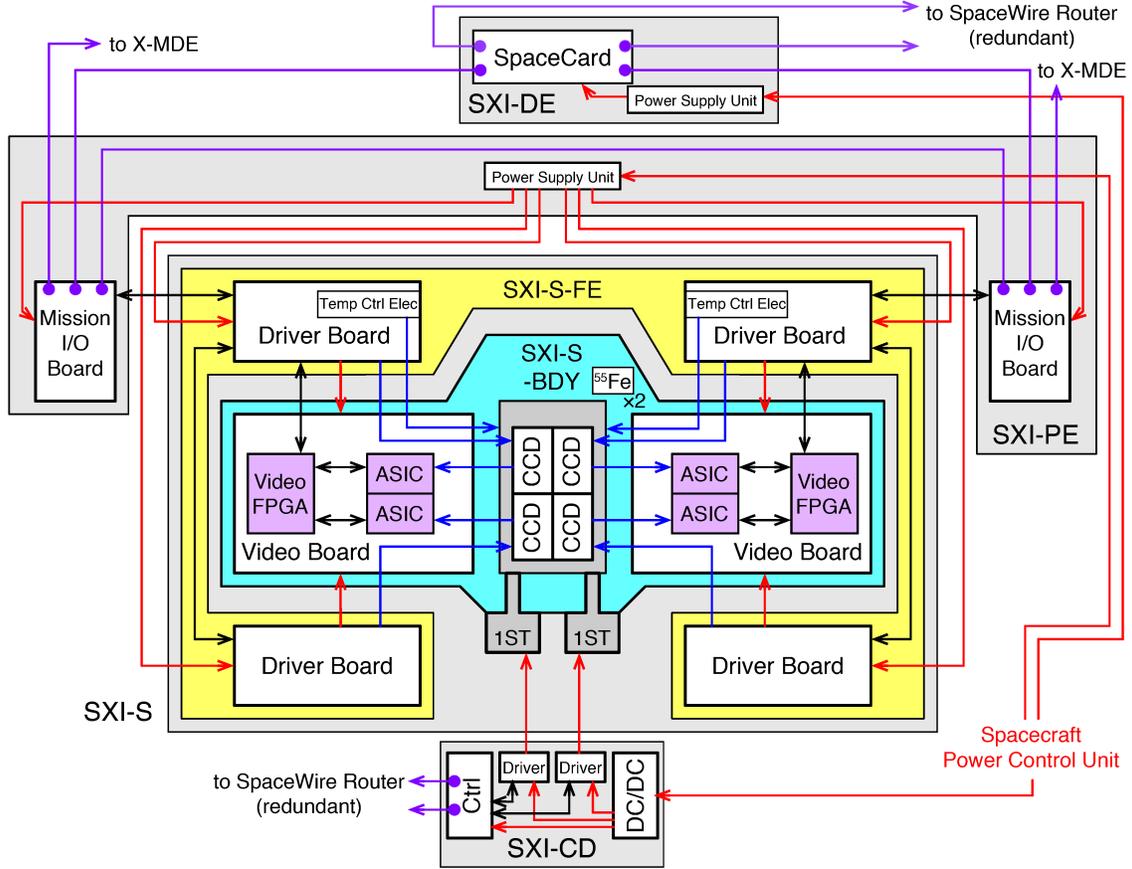}
\end{tabular}
\end{center}
\caption 
{ \label{fig:block_diagram}
Block diagram of the SXI system. The red, blue, purple, and black lines indicate power, analog, SpaceWire, and the other digital lines, respectively.} 
\end{figure}

Figure~\ref{fig:block_diagram} is a block diagram of the SXI system. The SXI consists of the sensor part (SXI-S) which includes CCDs, 
analog electronics (the Video Boards and SXI-S-FE) and mechanical coolers (SXI-S-1ST), 
the pixel processing electronics (SXI-PE), the digital electronics (SXI-DE), and the cooler driver (SXI-CD).  
The spacecraft has a backup digital electronics box, X-MDE (eXtra Mission Digital Electronics). 
X-MDE can act as any of the digital electronics for the scientific instruments aboard the spacecraft, including SXI-DE, in case of a failure. 
The SXI-S is placed at the focus of SXT-I on the base panel of the spacecraft whereas the other components are installed on the side panels. 
The power for the SXI system is supplied by the Power Control Unit (PCU) of the spacecraft whose output voltage ranges from 32~V to 52~V. 
The voltage is then regulated inside SXI-PE, DE and CD. The whole SXI system consumes a power of $\sim 85$~W. 
The power lines to SXI-PE, DE and CD consume about $25$~W, $10$~W, and $50$~W, respectively.  
The SXI system communicates with the spacecraft bus based on the SpaceWire protocol\cite{spw2005}. SXI-PE, DE and CD are 
connected with spacecraft's SpaceWire network via routers. 

\section{SXI Sensor (SXI-S)}
\subsection{Camera Body}
\begin{figure}
\begin{center}
\begin{tabular}{c}
\includegraphics[width=12cm]{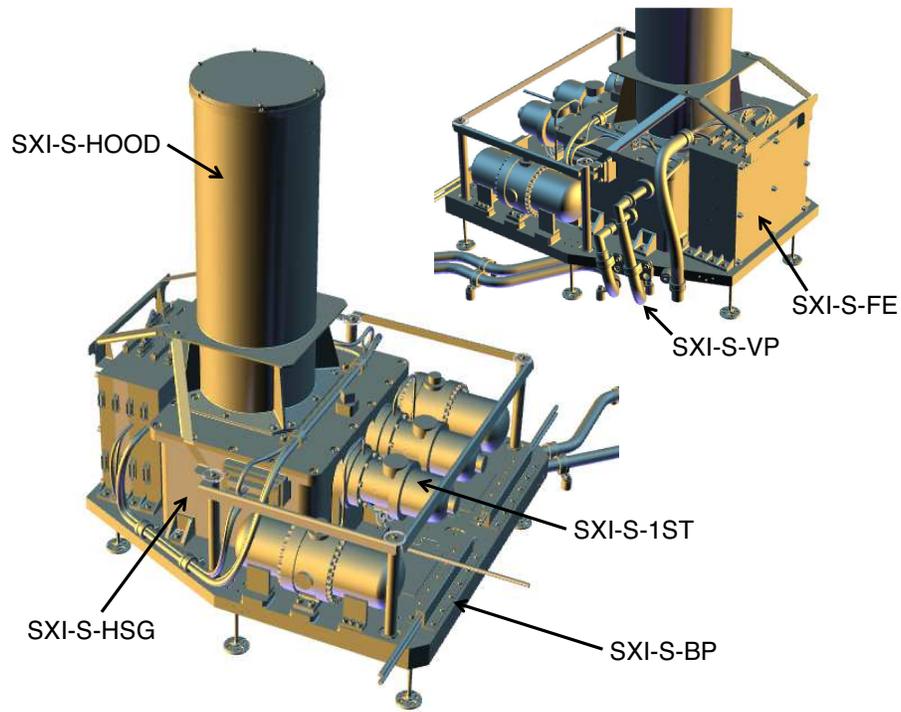}
\end{tabular}
\end{center}
\caption 
{ \label{fig:sxi-s_ext_view}
External views of SXI-S from different angles.} 
\end{figure}

\begin{figure}
\begin{center}
\begin{tabular}{c}
\includegraphics[width=6cm]{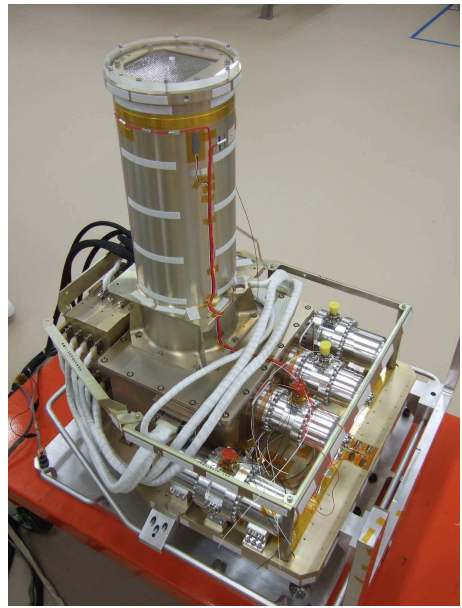}
\end{tabular}
\end{center}
\caption 
{ \label{fig:sxi-s_photo}
Photograph of the actual flight model of SXI-S. Note that the plastic cover attached on top of the hood is a non-flight item to protect the CBF during on-ground tests.} 
\end{figure}

We show external views and a photograph of SXI-S in Figs.~\ref{fig:sxi-s_ext_view} and \ref{fig:sxi-s_photo}, respectively. 
The camera body (SXI-S-BDY), made of Al alloy, consists of a hood (SXI-S-HOOD), a bonnet (SXI-S-BNT), a camera housing (SXI-S-HSG), and a base plate 
(SXI-S-BP). Inside the hood, there are five baffles to block stray light X-rays. 
SXI-S-BP is mounted $10~{\rm cm}$ above the base panel of 
the spacecraft with seven shafts made of Ti alloy. 
The shafts are designed so that the thermal expansion or contraction of SXI-S does not distort the base panel of the spacecraft, which is critical for the alignment.
Also, the design helps to reduce heat exchange between SXI-S and the spacecraft  via conduction. 
The whole camera body is covered by multi-layer insulators to shut out heat exchange with spacecraft via radiation. 
Heat generated in SXI-S is transferred to radiators by two heat pipes attached to SXI-S-BP. 

The camera body works as a radiation shield for the CCDs. 
Radiation damage results in increase of dark current and of CTI of the CCDs. 
In the case of a satellite in a low-earth orbit such as Hitomi, most of the radiation damage is due to protons trapped in the South Atlantic Anomaly (SAA). 
In order to stop SAA protons, we need to have a shield of $\gtrsim 10~{\rm g}~{\rm cm}^{-2}$ corresponding to 
an Al thickness of $\gtrsim 37~{\rm mm}$. 
Although a thicker body works better for shielding CCDs against SAA protons and also against unfocused X-rays, 
it results in higher instrumental backgrounds due to secondary particles. 
We did an optimization study based on Monte Carlo simulations\cite{tsuru2006,murakami2006} and chose to have a shield thickness of $35~{\rm mm}$ for most parts of the camera body. 
The surface of the inner side of the camera body is coated with Au over Ni plating to block Al-K fluorescence lines induced by cosmic rays. 
The Au layer has a thickness of $2~\mu{\rm m}$ to sufficiently absorb fluorescence lines from the underlying Ni layer.  

Four CCDs are installed on the cold plate made of Au-coated AlN inside the camera body.  
The cold plate is supported by six posts made of Torlon\textregistered{} polyamide-imide to achieve good thermal insulation from 
the camera body. Two heaters are glued to the cold plate for temperature control. 
The Video Boards, which digitize signals output by the CCDs, are also 
installed inside the body for better noise performance. 
In order to prevent contamination of the X-ray incident surface of the CCDs  
by outgas from the Video Boards, the space inside the camera body is physically 
divided into two: the upper room for the CCDs and the lower room for the Video Boards. 
Each room is directly connected to outside the spacecraft through vent pipes (SXI-S-VP).

\subsection{CCD}
\begin{figure}
\begin{center}
\begin{tabular}{c}
\includegraphics[width=9cm]{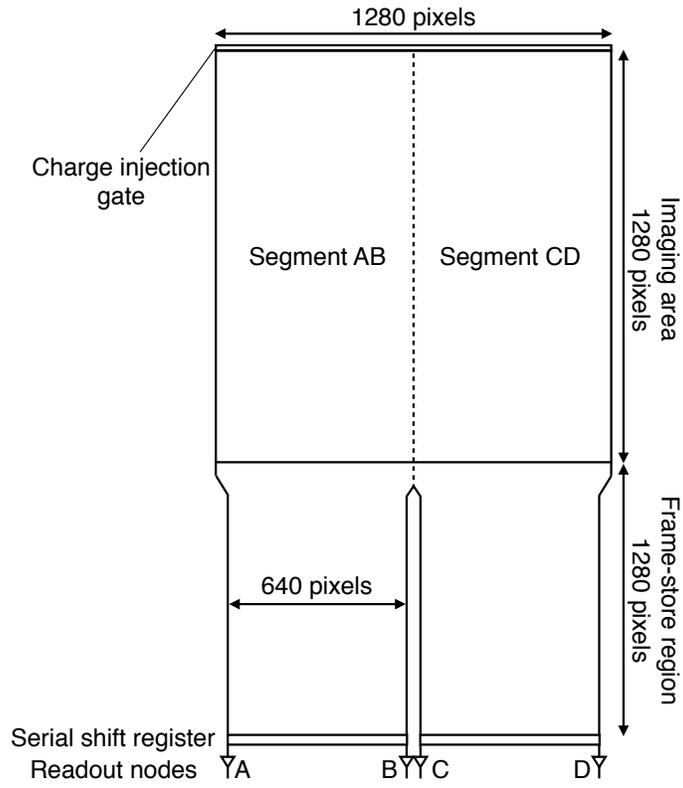}
\end{tabular}
\end{center}
\caption 
{ \label{fig:ccd_schematics}
Schematic view of the CCD, Pch-NeXT4. The segment AB (CD) are read out either from the readout node A or B (C or D).} 
\end{figure}

\begin{figure}
\begin{center}
\begin{tabular}{c}
\includegraphics[width=8cm]{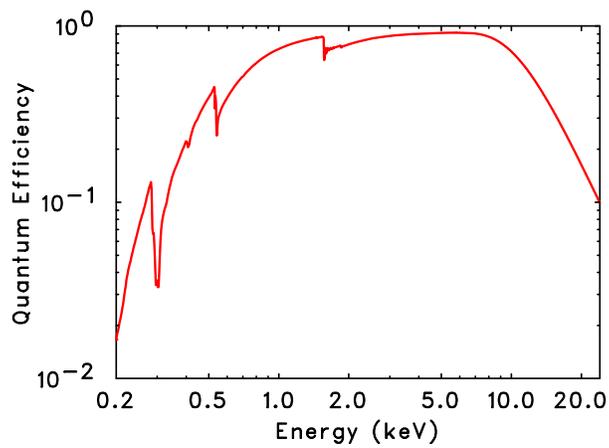}
\end{tabular}
\end{center}
\caption 
{ \label{fig:qe}
QE of the SXI as a function of X-ray energy. Absorption by the CBF and OBL is taken into account.} 
\end{figure}

The sensors used in the SXI are fully-depleted, back-illuminated CCDs named Pch-NeXT4 that we developed with Hamamatsu Photonics K.K.\cite{matsuure2006,ozawa2006,takagi2006,ueda2011,ueda2013} 
We summarize the specifications of Pch-NeXT4 in Tab.~\ref{tab:ccd_spec} and present its schematic view in Fig.~\ref{fig:ccd_schematics}. 
Pch-NeXT4 is the first P-channel CCD utilized for X-ray astronomy. 
The development was done also in collaboration with the National Astronomical Observatory of Japan, which was developing CCDs with similar specifications for the Hyper Suprime-Cam (HSC) of the Subaru Telescope\cite{hsc}.  
We applied a different treatment to the incident surface of the device from that used for the HSC CCDs to 
achieve better responses to soft X-rays\cite{ueda2013}.
Quantum efficiency (QE) of the CCD is plotted as a function of incident X-ray energy in Fig.~\ref{fig:qe}. 
The n-type substrate with high resistivity enables a thick ($200~\mu{\rm m}$) depletion layer, resulting in high QE for the hard X-ray band whereas 
the back-illumination (BI) configuration enhances QE in the soft X-ray band. 
The BI CCD is also advantageous for high resistance to micrometeoroid damage\cite{stuhlinger2006}. 

The imaging area of Pch-NeXT4 has a  $24~\mu{\rm m} \times 24~\mu{\rm m}$ pixel size and $1280 \times 1280$ pixels, which 
yields an imaging area size of $30.72~{\rm mm} \times 30.72~{\rm mm}$. 
Since we nominally apply on-chip $2 \times 2$ binning, the effective pixel size and pixel format are  $48~\mu{\rm m} \times 48~\mu{\rm m}$ and 
$640 \times 640$, respectively. 
In what follows, we refer to the $24~\mu{\rm m} \times 24~\mu{\rm m}$ and $48~\mu{\rm m} \times 48~\mu{\rm m}$ pixels
as physical and logical pixels, respectively. 
The Suzaku XIS team demonstrated that the CTI increased by radiation damage can be restored by using a charge injection technique\cite{uchiyama2009}. 
Pch-NeXT4 also has charge-injection capability\cite{mori2013}. We can inject artificial charges from the gates attached to the top of each column. 
We inject charges into every 160th physical row in each frame. 
The CTI could be restored more if we inject charges to more rows. Charge injection rows, however, cannot be used in observations. 
The frequency of charge injection rows was determined based on the balance between the CTI restoration and the loss of the effective rows. 
The amount of injected charges is controlled so that signals from the charge injection rows are saturated but charge leakage to leading and trailing rows is negligible. 
In order to block optical light, the incident surface of the CCD is coated with the Optical Blocking Layer (OBL)\cite{kohmura2010}, a 100-nm thick Al layer 
over a 20-nm thick ${\rm SiO}_2$ layer. 
The optical light transmission of the OBL is in the order of $10^{-5}$.  
Each CCD has four readout nodes (Fig.~\ref{fig:ccd_schematics}). 
We read out a half of the imaging area (a segment) from one of the nodes. 
We nominally read out signals from the nodes A and C and use the nodes B and D as a redundant option. 

\begin{figure}
\begin{center}
\begin{tabular}{c}
\includegraphics[width=9cm]{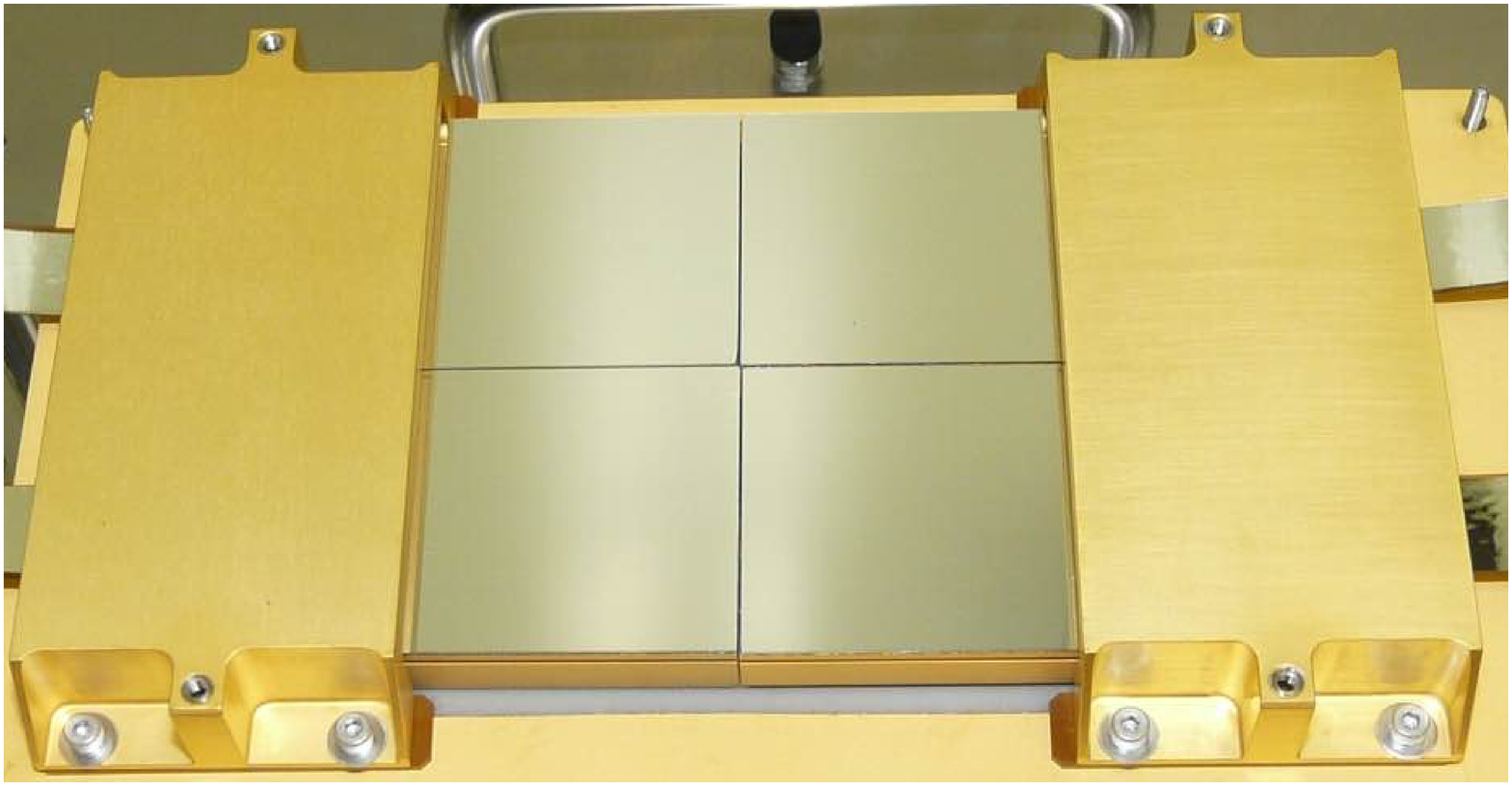}
\end{tabular}
\end{center}
\caption 
{ \label{fig:ccd_photo}
Photograph of the flight CCDs mounted on the cold plate.} 
\end{figure}

\begin{figure}
\begin{center}
\begin{tabular}{c}
\includegraphics[width=9cm]{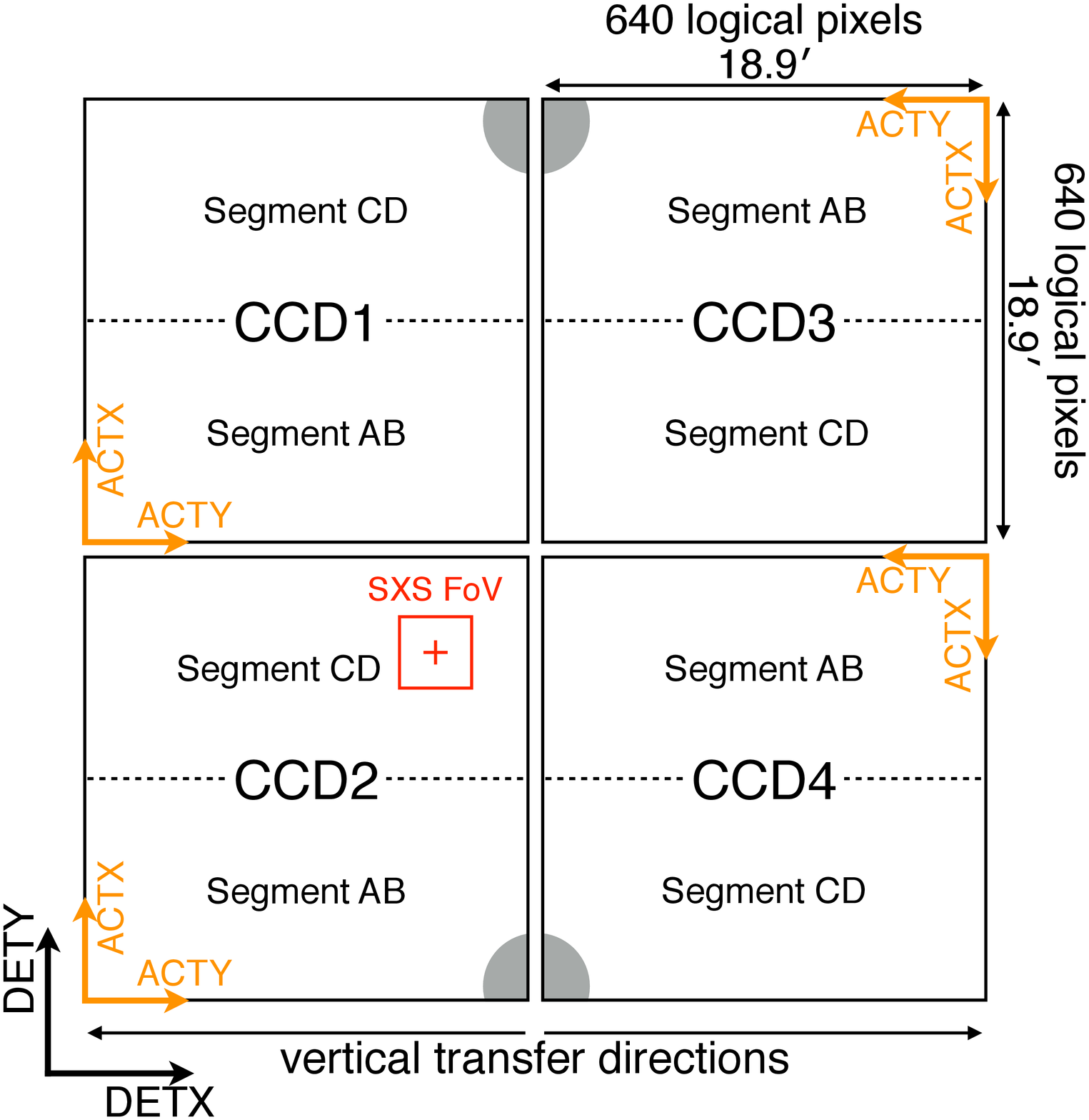}
\end{tabular}
\end{center}
\caption 
{ \label{fig:ccd_layout}
Schematic layout of the CCD arrays in a look-up view. Only the imaging areas are drawn. The red cross and square indicate the aim point of SXT-I and the 
FoV of the SXS, respectively. The gray shaded zones are approximate locations illuminated by the $^{55}{\rm Fe}$ calibration sources. The definitions of the DET and ACT coordinates are indicated in black and orange, respectively.} 
\end{figure}

We show a photograph of the actual flight CCDs in Fig.~\ref{fig:ccd_photo} and a schematic layout of the CCD array in Fig.~\ref{fig:ccd_layout}. 
The Pch-NeXT4 chips are arranged in a  $2 \times 2$ array. The typical spacing between the chips is $\sim 700~{\mu}{\rm m}$ according to the measurement described in \S\ref{subsec:mesh}. 
Since the CCD packages are installed directly on the flat cold plate, they are almost co-planer and their tilts are negligible. 
The aim point of the SXT is offset from the array center to prevent targets from falling into the gap between the CCDs. 
Combined with the SXT, the imaging areas correspond to a $38^{\prime} \times 38^{\prime}$ FoV. 
The frame-store regions are covered with 3-mm thick Al shields coated with Au to block focused X-rays. 
The CCD signals are carried to the Video Boards through flexible printed circuits. 
For gain monitoring, two $^{55}{\rm Fe}$ sources are installed into the bonnet and illuminate a corner of each 
CCD.

\begin{table}[ht]
\caption{Specifications of the SXI CCD, Pch-NeXT4.} 
\label{tab:ccd_spec}
\begin{center}       
\begin{tabular}{cc} 
\hline
\hline
\rule[-1ex]{0pt}{3.5ex}  Architecture & Frame Transfer  \\
\rule[-1ex]{0pt}{3.5ex}  Channel Type & P-channel   \\
\rule[-1ex]{0pt}{3.5ex}  Pixel Format & $1280 \times 1280$  \\
\rule[-1ex]{0pt}{3.5ex}  Imaging Area Size &  $30.720~{\rm mm} \times 30.720~{\rm mm}$ \\
\rule[-1ex]{0pt}{3.5ex}  Chip Size &  $31.220~{\rm mm} \times 57.525~{\rm mm}$ \\
\rule[-1ex]{0pt}{3.5ex}  Pixel Size (Imaging Area)  & $24~\mu{\rm m} \times 24~\mu{\rm m}$ \\
\rule[-1ex]{0pt}{3.5ex}  Pixel Size (Frame-store Region)  & $22$--$24~\mu{\rm m}~({\rm H}) \times 16~\mu{\rm m}~({\rm V})$ \\
\rule[-1ex]{0pt}{3.5ex}  Depletion Layer Thickness & $200~\mu{\rm m}$  \\
\rule[-1ex]{0pt}{3.5ex}  Clock Phase & 2   \\
\rule[-1ex]{0pt}{3.5ex}  Incident Surface Coating & $100$-${\rm nm}$-thick Al  \\
\rule[-1ex]{0pt}{3.5ex}  Output & 1-stage MOSFET Source Follower \\
\rule[-1ex]{0pt}{3.5ex}  Node Sensitivity & $5~\mu{\rm V}/e^{-}$\\
\rule[-1ex]{0pt}{3.5ex}  Frame Cycle & $4~{\rm s}$\\
\rule[-1ex]{0pt}{3.5ex}  Pixel Rate & $69.44~{\rm kHz}$\\
\rule[-1ex]{0pt}{3.5ex}  Readout Noise (CCD) & $4$--$5~e^{-}$ (rms)\\
\rule[-1ex]{0pt}{3.5ex}  Readout Noise (Whole SXI System) & $6$--$7~e^{-}$ (rms)\\
\hline
\end{tabular}
\end{center}
\end{table} 

\subsection{Video Board}
The primary function of the Video Board is analog-to-digital conversion of CCD signals by $\Delta\Sigma$ modulation. 
The main components of the Video Board are four analog ASICs, MND02\cite{MND02,nakajima2013}, and a field-programmable gate array (FPGA), RTAX2000 by Microsemi. 
The MND02 chip has four channels and each channel contains a preamplifier, a 5-bit digital-to-analog converter (DAC), and 
two $\Delta\Sigma$ modulators. 
The gain of the preamplifier is adjustable between 0.6 and 10 in nine steps. 
Pulse height is obtained by subtracting a floating level voltage from a signal level voltage of a CCD signal. 
Even with no signal charges, the voltage difference becomes non-zero. 
The DAC is for canceling out this offset. 
In this way, we can make most of the input signal range of the $\Delta\Sigma$ modulators which 
convert an analog signal to a digital 155-bit stream. 
The ASIC has two modulators per channel, which are operated alternately to achieve a high pixel rate. 
The FPGA sends control signals to MND02, and converts the bit stream into 12-bit pulse height data 
by a decimation filter. 
The digitized frame data are then sent to SXI-PE for further processing. 
One board processes signals from two CCDs and thus the SXI has two boards in the camera body.  
Each of the readout nodes of the CCD is fed to a pair of channels of different ASICs, which ensures redundancy. 
Taking advantage of the connection, we nominally average the two outputs to improve the readout noise as we demonstrated 
with prototype hardware\cite{nakajima2009}.  
This method is effective since the noise generated in the ASICs is not negligible. 

\subsection{Front-end Electronics (FE)}
SXI-S-FE is composed of four CCD Driver Boards, each of which is connected to one of the four CCDs, and
is mounted on SXI-S-BP next to the camera housing (Fig.~\ref{fig:sxi-s_ext_view}). 
The driver boards generate analog clocks and biases for the CCDs and supply them to the CCDs 
via the Video Boards.  
The back-bias voltage for the CCD substrate, for which we typically apply $35~{\rm V}$, is also generated in 
SXI-S-FE using a Cockroft-Walton charge pump voltage multiplier. 
The high and low level voltages of the clocks are determined by outputs from on-board DACs and 
the two levels are switched according to digital timing signals sent from SXI-PE. 
The voltage levels for the bias lines are also controlled by DACs. 
SXI-S-FE measures voltage levels of the clock and bias lines and outputs them to SXI-PE as housekeeping (HK) data 
after analog-to-digital conversion and multiplexing. 
Two of the four Driver Boards also take responsibility as temperature control electronics (TCE) for the CCDs (Fig.~\ref{fig:block_diagram}). 
Each of them measures temperatures of two CCDs and outputs current to one of the heaters attached to the cold plate. 

\subsection{Cooling System}
Single-stage Stirling coolers, SXI-S-1ST, are used to cool the CCDs to the operation temperature. 
The SXI has two Stirling coolers. Only one of them is operated in orbit, and the other 
is for standby redundancy. 
The cooler consists of a cold head, a compressor, and a capillary tube to connect them. 
The cold heads and compressors are installed into SXI-S-HSG and on SXI-S-BP, respectively. 
Active balancers are inside the cold heads to reduce vibration induced along the drive axis. 
The cold finger of each cooler is connected to the cold plate with a Cu flexure structure. 
This design ensures higher thermal conductivity 
while less vibration is transferred from the cold fingers to the cold plate and 
the difference of thermal expansion can be absorbed. 
SXI-S-1ST is controlled and monitored independently of the camera by SXI-CD.  
In nominal operations, we supply a constant power to SXI-S-1ST and stabilize the CCD temperature 
at a targeted temperature by changing the heater current  with a propotional-integral-derivative (PID) controller implemented in the 
on-board software of SXI-DE (\S\ref{subsec:temp_ctrl}). 

\subsection{Contamination Blocking Filter (CBF)}
The low-energy QE of the XIS aboard the Suzaku satellite decreased after the launch because of 
the accumulation of contaminating material on the optical blocking filter\cite{XIS}.
Since the filter was placed close to the cold CCD surface, the filter was colder than the other parts of 
the spacecraft and thus adsorbed the contaminant. 
This experience led us to the design of the SXI in which the CBF\cite{kohmura2014} is placed on top of the hood. 
The CBF is a 200-nm thick polyimide film with a stainless steel mesh support. 
Al is vapor-deposited on both sides of the film. 
The temperature of the CBF is kept at $\sim 25~^\circ{\rm C}$ by heaters attached to SXI-S-HOOD.

\section{SXI Pixel Process Electronics (SXI-PE)}
SXI-PE consists of a Power Supply Unit (PSU) and two Mission I/O (MIO) boards (Fig.~\ref{fig:block_diagram}). 
The PSU generates DC voltages for SXI-S and the MIO boards from the spacecraft's bus power line. 
The MIO boards are developed to be commonly used for all the scientific instruments aboard Hitomi. 
Each of the MIO boards has two FPGAs (RTAX2000) and an SDRAM. One of the FPGAs is called SpaceWire FPGA, 
which provides an interface for SpaceWire communication and also an interface for the SDRAM. 
A user-specific logic is implemented on the other FPGA, UserFPGA. 
One MIO board takes care of two CCDs. One of the MIO boards can provide clock signals to the other 
board to avoid possible interferences. 
The functions of SXI-PE are CCD clock pattern generation, CCD data processing, autonomous HK data 
collection from SXI-S-FE, and control of the DACs of SXI-S-FE in response to directions by SXI-DE.

\subsection{CCD Clock Pattern Generation}
SXI-PE provides SXI-S-FE with digital timing signals, based on which analog CCD clocks are generated. 
A microcode program loaded to SXI-PE determines the clock pattern and a four-bit pixel code for each pixel read out. 
The pixel code specifies the attributes of the pixel, for example, if it is an imaging area pixel or is an under/over-clock pixel. 
The architecture of the microcode program is similar to that used for Suzaku XIS. 
The timing signals are synchronized with a 10~MHz clock and thus we can assign the logical levels of each 
signal in every $0.1~\mu{\rm s}$. In all the clocking modes described below, the vertical transfer time is 
$28.8~\mu{\rm s}$ per physical pixel row for the transfer from the imaging area to the frame-store region (hereafter fast transfer), 
and is $5.24~{\rm ms}$ per logical pixel row for the transfer in the frame-store region (hereafter slow transfer).
The readout time is $14.4~\mu{\rm s}$ per logical pixel. 

\begin{figure}
\begin{center}
\begin{tabular}{c}
\includegraphics[width=16cm]{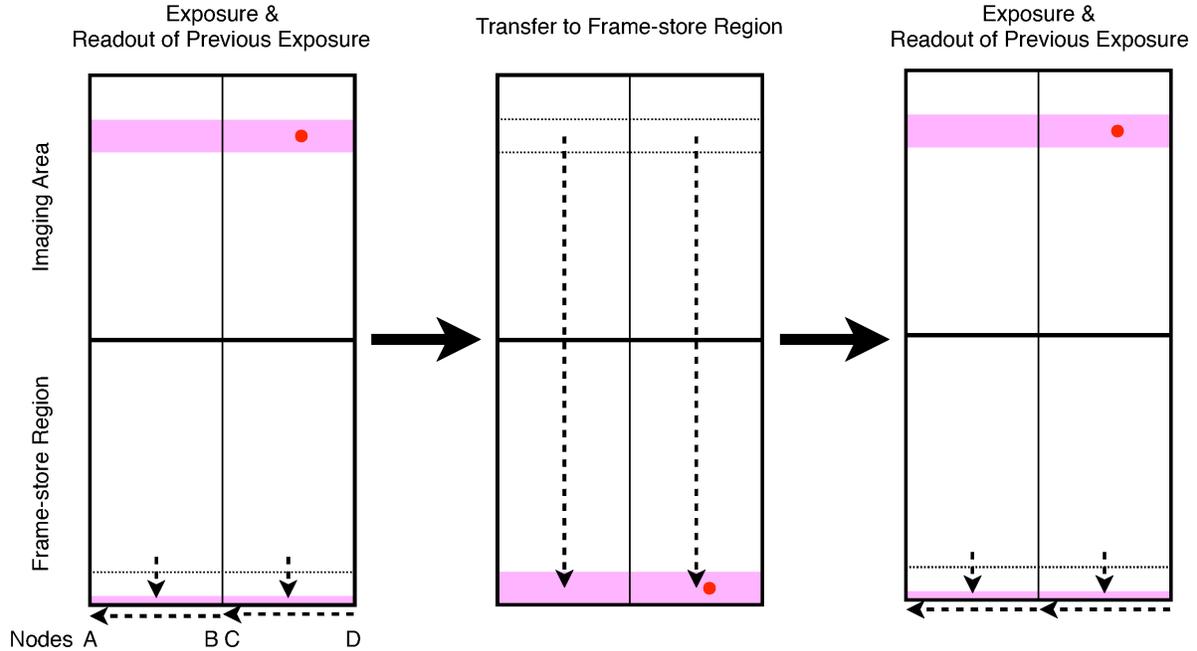}
\end{tabular}
\end{center}
\caption 
{\label{fig:window_option}
Schematic diagram of the window option operation.} 
\end{figure}

\begin{figure}
\begin{center}
\begin{tabular}{c}
\includegraphics[width=16cm]{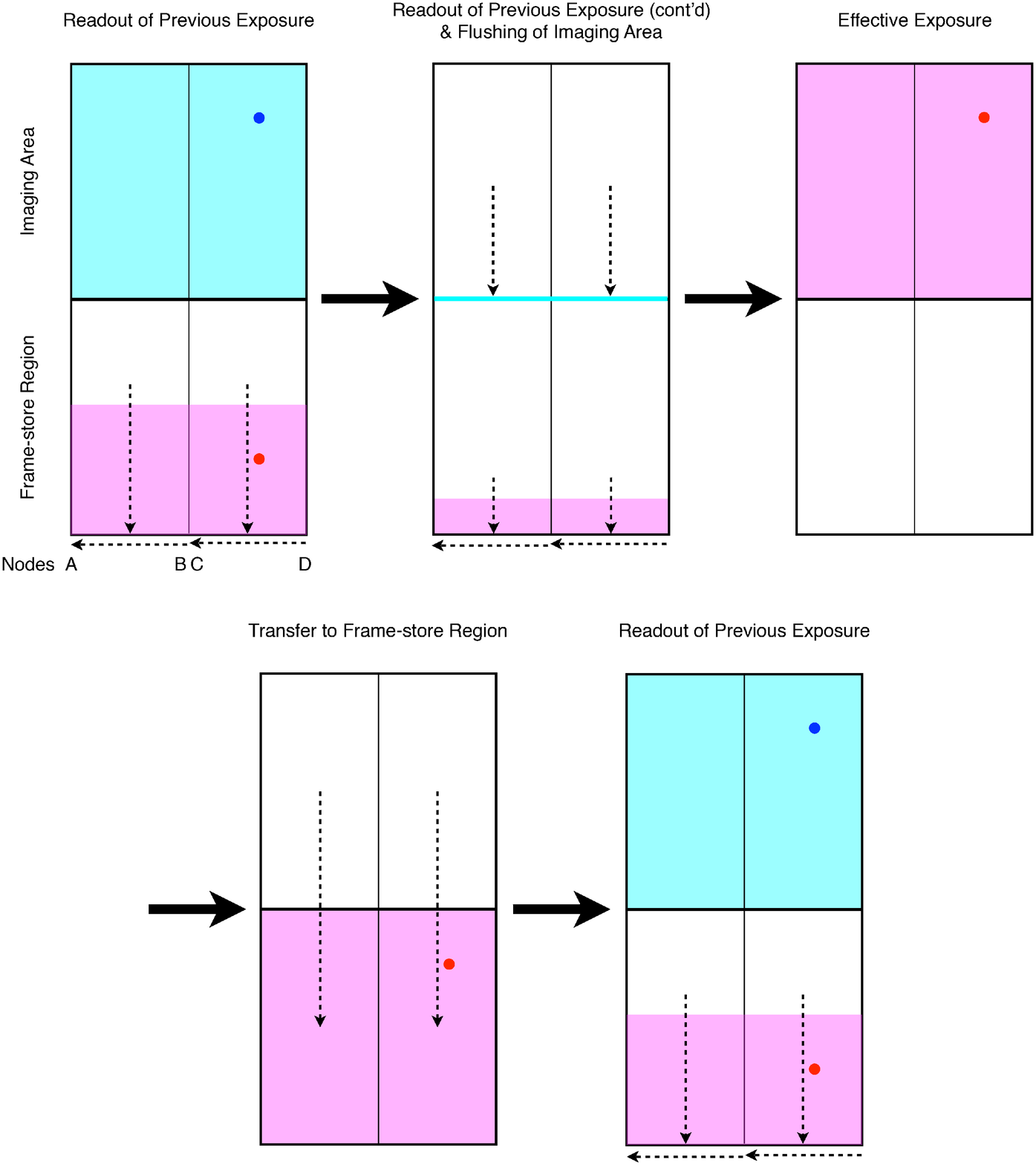}
\end{tabular}
\end{center}
\caption 
{\label{fig:burst_option}
Same as Fig.~\ref{fig:window_option} but for the burst option operation.} 
\end{figure}

Listed in Tab.~\ref{tab:clocking_modes} are the clocking modes we planned to support for in-orbit observations. 
When observing bright sources, we can reduce the probability of pile-up by a window option,  
a burst option, and their combination. 
Schematic drawings of the window option and burst option are presented in Figs.~\ref{fig:window_option} and \ref{fig:burst_option}. 
In the window option, a limited number of rows of the imaging area are read out, which results in shorter and 
frequent readout. 
We read out 80 out of 640 logical pixel rows with the 1/ 8 window option.
Although the rows outside the specified window are also transferred to the frame-store region, they are 
binned on-chip into a single row in the output serial register and then this row is flushed and ignored. 
The burst option is for partial readout in time, and only a small fraction of the frame cycle is used as the effective exposure time. 
In the burst option with an exposure time of $t_{\rm exp}$, the CCD is first exposed for a duration of $(4~{\rm s} - t_{\rm exp})$. 
The CCD imaging area is cleared of charges accumulated during this period by transferring them to the bottom row of the 
imaging area. 
Then, an effective exposure of $t_{\rm exp}$ is started. 
After this exposure, accumulated charges are transferred to the frame-store region, and are read out in the same manner 
as in the clocking modes without the burst option. 
We support two burst options with $t_{\rm exp} = 1.9396~{\rm s}$ and $t_{\rm exp} = 0.0606 ~{\rm s}$. 

\begin{table}[ht]
\caption{Clocking modes of the SXI.} 
\label{tab:clocking_modes}
\begin{center}       
\begin{tabular}{cccc} 
\hline
\hline
\rule[-1ex]{0pt}{3.5ex}  Option Name & Logical Pixels & Exposure Time &  Exposure per Frame    \\
\rule[-1ex]{0pt}{3.5ex}        & (${\rm H} \times {\rm V}$) & [s] & \\
\hline
\rule[-1ex]{0pt}{3.5ex}  Full Window + No Burst & $640 \times 640$  &  3.9631 & 1 \\
\rule[-1ex]{0pt}{3.5ex}  Full Window + 2-s Burst & $640 \times 640$  & 1.9396 & 1\\
\rule[-1ex]{0pt}{3.5ex}  Full Window + 0.1-s Burst & $640 \times 640$  &  0.0606 & 1\\
\rule[-1ex]{0pt}{3.5ex}  1/8 Window + No Burst & $640 \times 80$  &  0.4631 & 8\\
\rule[-1ex]{0pt}{3.5ex}  1/8 Window + 0.1-s Burst & $640 \times 80$  &  0.0606 & 8\\
\hline
\end{tabular}
\end{center}
\end{table} 

\subsection{CCD Data Processing}
Another major role of SXI-PE is processing CCD signals.  
SXI-PE receives raw, digitized data from the Video Boards of SXI-S. 
According to settings written in registers, outputs from designated channels of the ASICs are selected and 
averages of signals from the same CCD readout nodes are calculated. 
Then, the pixel codes 
are appended to the data from each pixel by referring to information written in the microcode. 

Dark level is estimated and subtracted from the raw frame data before further data processing. 
Dark levels are different for each pixel and can be also time-variable. 
SXI-PE, therefore, calculates dark levels for each individual pixel and updates them for every frame. 
If dark-subtracted pulse height (hereafter PH) is between the lower and upper thresholds for dark update, 
the dark level (${\rm Dark}$) is replaced by ${\rm Dark} + {\rm PH}/{h}$, where $h$ is a parameter 
called ``History Parameter'' whose default and nominally used value is 8. 
This parameter determines how promptly the dark level estimation reacts to the changes of PH. 
This algorithm, which was comprehensively verified with prototype hardware, 
 allows us not only to improve accuracy of the dark estimation as the frame cycle continues but also 
to follow gradual changes of the dark level with time. 
Pixels are flagged as hot pixels if their dark levels are higher than a preset threshold. 
The dark calculation and the hot pixel list can be initialized by sending a command. 
We planned to do so at the beginning of each observation. 

After the dark level subtraction, the UserFPGA program searches for X-ray event candidates in 
the one-dimensional pixel data stream. 
If PH of a pixel is between the lower and upper thresholds for event detection and the PH is higher than that of the adjacent left pixel and higher than or 
equal to that of the adjacent right pixel, the pixel is regarded as the event candidate center pixel. 
The event candidate list as well as the frame data is then sent to SXI-DE, where more detailed data processing is performed. 
Although only a crude extraction of event candidates is possible with this simple logic implemented in SXI-PE, 
this makes the data processing later in SXI-DE efficient.

\section{SXI Digital Electronics (SXI-DE)}
SXI-DE has a CPU board called SpaceCard and a PSU to supply DC power to it (Fig.~\ref{fig:block_diagram}). 
Similarly to the MIO boards, the SpaceCard boards are commonly used for all the scientific instruments aboard the spacecraft. 
SXI-DE interfaces the SXI system with the spacecraft's bus system, receiving commands from the Satellite Management Unit (SMU) and 
sending telemetry such as event data and HK data to the Data Recorder (DR) and the SMU. 
Also, SXI-DE performs further CCD data processing and monitor/control the whole SXI system except for the Stirling coolers, SXI-S-1ST, which 
is under the control of SXI-CD. 
In what follows, we will describe the CCD data processing and how SXI-DE controls the current through the heaters attached to the 
cold plate for CCD temperature control.  

\subsection{CCD Data Processing}
Firstly, the average PH of horizontal overclocked (HOC) pixels of the row is subtracted from PH from imaging area pixels so that 
we can remove the effect of short-time variability which cannot be followed by the dark level calculation. 
Then, three kinds of filters, area discrimination, surround filtering, and $3\times3$ local maximum filtering, are applied 
to further limit the amount of data. 
The area discrimination restricts imaging area regions for X-ray event search. 
The regions are defined as rectangles in the imaging area, and pixels either inside or outside the regions 
are excluded from the event identification process.  
For example, a very bright source in the FoV can be masked to avoid telemetry saturation. 
The surrounding filtering is for discarding charged particle background events in which generated charges are shared by 
a number of pixels along the track of the particles. 
This filter examines PH from the eight pixels surrounding the event center pixel. If the number of pixels whose PH exceeds 
a threshold is more than a preset number, the event is removed from the event list. The threshold and the number of pixels 
can be set by a command. 
The $3\times3$ local maximum filtering extracts events whose center pixel in the $3\times3$ pixel island has PH larger than 
the other eight pixels. 
This filtering is necessary since the SXI-PE data processing only compares PH of $3\times1$  pixels in the same row, and thus 
the same event may be double counted in adjacent rows. 

Event candidates which passed through all the filters are sent to the DR and then to ground stations as event data. 
The data include information on the location of the event and PH of $5 \times 5$ pixels. 
Here data from inner $3\times3$ pixels and data from outer 16 pixels of the $5 \times 5$ pixel island are divided 
into two separate packets. 
Since the latter packets have lower priority, they may not be sent to the ground. 
Therefore, $3\times3$ data include hit patterns of the outer 16 pixels indicating which pixels have 
PH larger than a threshold. 
Both $3\times3$ and $5 \times 5$ event data are compressed in SXI-DE to reduce the telemetry size.

\subsection{Heater Current Control}\label{subsec:temp_ctrl}
\begin{figure}
\begin{center}
\begin{tabular}{c}
\includegraphics[width=10cm]{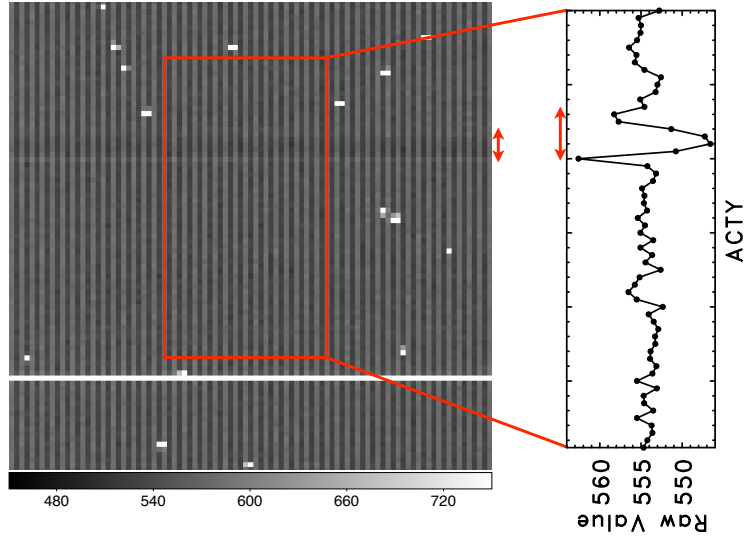}
\end{tabular}
\end{center}
\caption 
{\label{fig:downstream} 
Raw frame data from a $\sim 100 \times 100$ logical pixel region of CCD3 Segment AB taken in an on-ground test (left), and projection of the frame along the ACTY direction (right). 
The white row in the frame image is a charge injection row. 
Pixel values of the rows indicated with the red arrows becomes significantly higher/lower than those of the other rows. This phenomenon was found to be synchronized with a rapid change of the heater current.} 
\end{figure}

\begin{figure}
\begin{center}
\begin{tabular}{c}
\includegraphics[width=16cm]{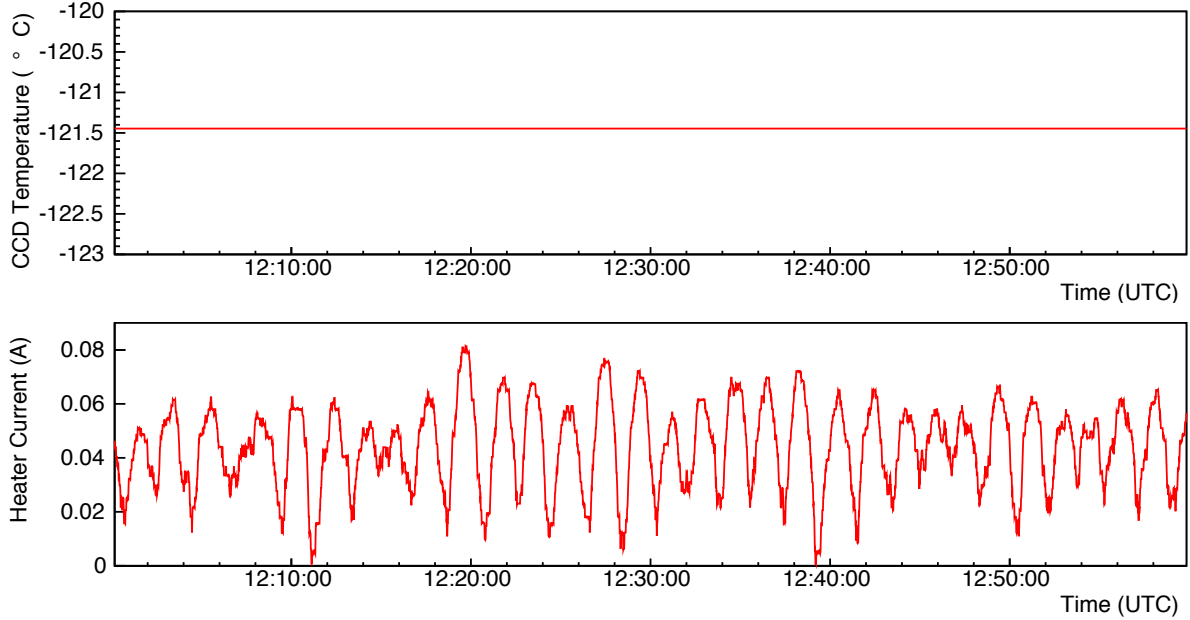}
\end{tabular}
\end{center}
\caption 
{\label{fig:ccd_temp} Time history of the CCD temperature and the heater current obtained in an on-ground test. 
The targeted temperature was set to $-121.4~^\circ{\rm C}$.  
The measured temperature was actually stabilized at the targeted temperature within one digit of the data which corresponds to $\sim 0.1~^\circ{\rm C}$. 
} 
\end{figure}

To keep the CCD temperature constant, the SXI-DE software controls 
the heater attached to the cold plate on the basis of the PID algorithm. 
With a frequency of 1~Hz, the CCD temperature is sampled and the heater current is adjusted to keep 
the temperature stabilized at a targeted value. 
If we define $P_i$ and $\Delta T_i$ as the power dissipated in the heater and 
the difference between the targeted temperature and current temperature after the $i$-th iteration, respectively, 
and $f$ as the frequency of the iteration, 1~Hz,  
the PID algorithm yields the equation 
\begin{eqnarray}
P_n - P_{n-1} = K_p \, (\Delta T_n -  \Delta T_{n-1}) + \frac{K_i \,  \Delta T_n}{f}  + K_d \, f [(\Delta T_n -  \Delta T_{n-1})-(\Delta T_{n-1} -  \Delta T_{n-2})], 
\end{eqnarray}
where $K_p$, $K_i$, and $K_d$ are constants called the proportional gain, the integral gain, and the derivative gain, respectively. 
In on-ground tests, we found an interference between the heater control line and CCD readout electronics.
When the heater current was decreased or increased as rapid as $\gtrsim 0.05~{\rm A}~{\rm s}^{-1}$, anomalies of PH appeared as show in Fig.~\ref{fig:downstream}. 
We tuned the parameters to suppress the changing rate of the heater current so that we can avoid the interference. 
In Fig.~\ref{fig:ccd_temp}, we plot the CCD temperature and heater current as a function of time taken in an on-ground test after the parameter optimization. 

\section{On-ground Data Processing and Pre-launch Performance}
\subsection{Coordinate Assignment}\label{subsec:mesh}
\begin{figure}
\begin{center}
\begin{tabular}{c}
\includegraphics[width=15cm]{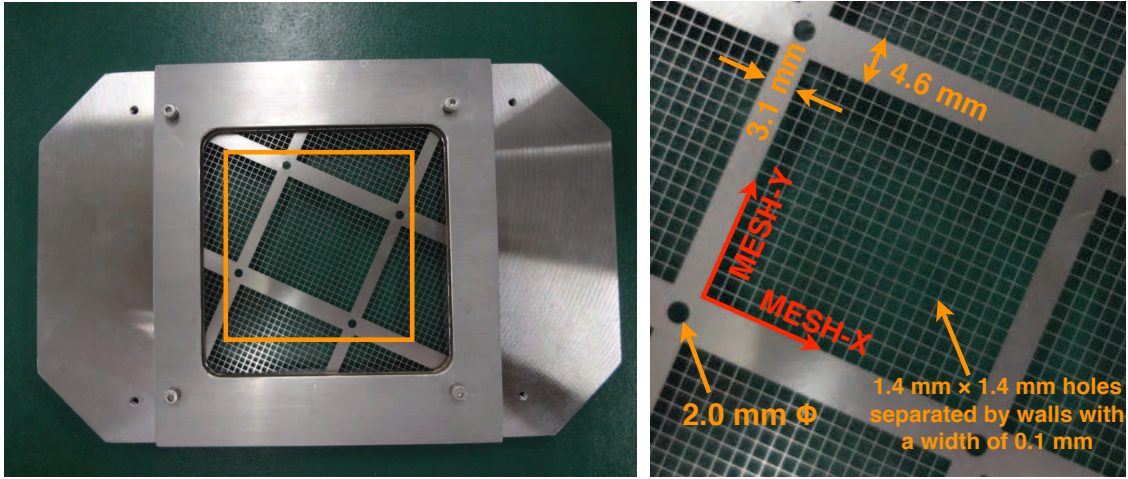}
\end{tabular}
\end{center}
\caption 
{\label{fig:mesh_photo}Photograph of the stainless steel mesh in a look-down view. The right photograph is a zoom-in view of the region enclosed by the orange square in the left photograph. 
The definition of the MESH coordinate is indicated in red.} 
\end{figure}

\begin{figure}
\begin{center}
\begin{tabular}{c}
\includegraphics[width=10cm]{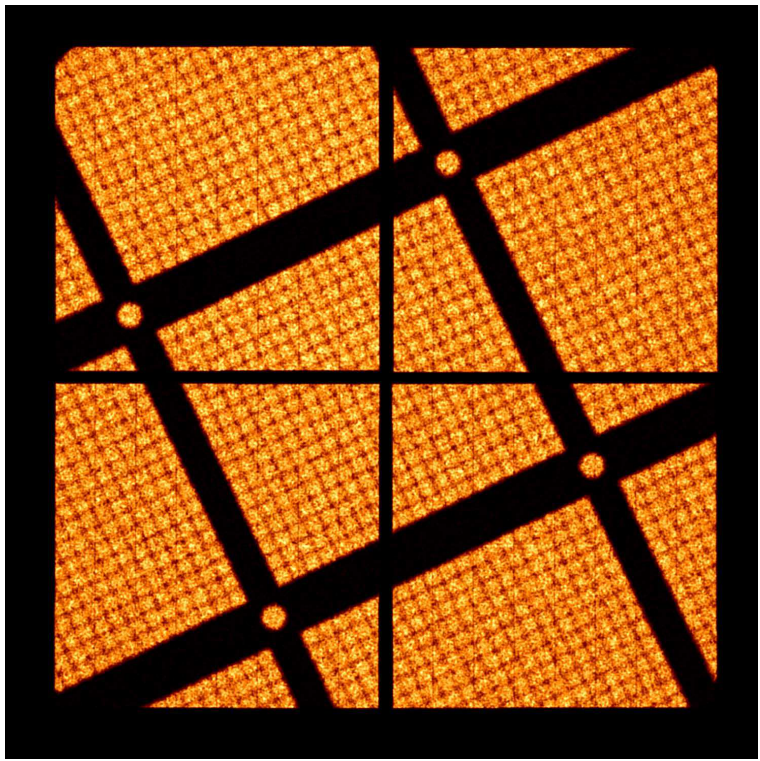}
\end{tabular}
\end{center}
\caption 
{\label{fig:mesh_image_det} 
X-ray shadow image of the mesh by the SXI CCD array in the DET coordinate defined in Fig.~\ref{fig:ccd_layout} (in a look-up view). 
We irradiated the CCDs with Mn-K$\alpha$ at 5.9~keV and K$\beta$ at 6.5 keV from 
an $^{55}{\rm Fe}$ source. The image is smoothed with a Gaussian kernel with $\sigma = 1.5$~logical pixels.} 
\end{figure}

A sky coordinate is assigned to each photon based on information on attitude of the spacecraft. 
In this process, we must accurately know the relative rotation angle and position of each CCD on the cold plate. 
We obtained the information by analyzing a shadow image of the mesh shown in Fig.~\ref{fig:mesh_photo}. 
The mesh is made of stainless steel and has a thickness of $0.1~{\rm mm}$. 
We placed the mesh $7~{\rm mm}$ above the surface of the CCDs and illuminated them with Mn-K$\alpha$ and K$\beta$ 
X-rays from an $^{55}{\rm Fe}$ source placed $202~{\rm mm}$ above the CCDs. 
In the analysis described below, we use two coordinate systems, the ACT and MESH coordinates, whose definitions are 
shown in Figs.~\ref{fig:ccd_layout} and \ref{fig:mesh_photo}, respectively. 
We first measured the rotation angle ($\theta$) of the ACT coordinate of each CCD with respect to the Mesh coordinate. 
We projected the shadow image along the MESH-X axis with various rotation angles assumed. 
The contrast of the projected distribution becomes the highest when the true angle $\theta$ is assumed. 
The angles $\theta$ were obtained to be ${25.40^\circ}^{+0.07^\circ}_{-0.06^\circ}$, $25.72^\circ \pm 0.06^\circ$, ${25.56^\circ}^{+0.10^\circ}_{-0.09^\circ}$, and $25.42^\circ \pm 0.09^\circ$ 
for CCD1, CCD2, CCD3, and CCD4, respectively. 
We then estimated the relative positions of the CCDs along the MESH-X and MESH-Y axes. 
We projected the counts map along the MESH-X and MESH-Y axes, and determined the positions of the CCDs with which the mesh pattern agrees with each other.  
Shown in Fig.~\ref{fig:mesh_image_det} is the resultant shadow image of the mesh in the DET coordinate (Fig.~\ref{fig:ccd_layout}), 
which demonstrates the imaging capability of the SXI. 

\subsection{Uniformity of Quantum Efficiency}
\begin{figure}
\begin{center}
\begin{tabular}{c}
\includegraphics[width=16cm]{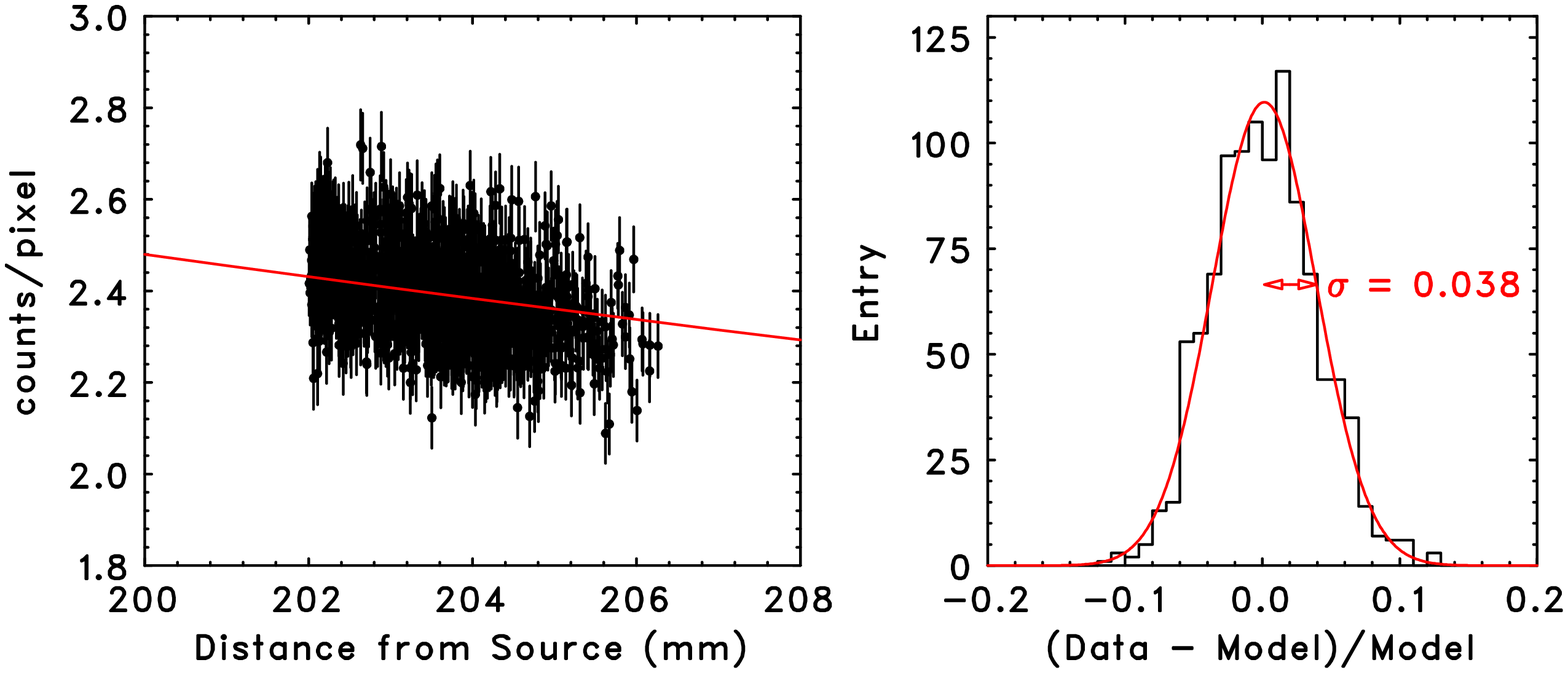}
\end{tabular}
\end{center}
\caption 
{\label{fig:mesh_qe} 
(left) Relation between X-ray counts and distance from the  $^{55}{\rm Fe}$ source, $R$, in the shadow image in Fig.~\ref{fig:mesh_image_det}. 
Each data point represents X-ray counts accumulated in regions corresponding to each of the $1.4~{\rm mm} \times 1.4~{\rm mm}$ holes of the mesh (Fig.~\ref{fig:mesh_photo}). 
The red curve indicates the function $A/R^2$, where the normalization $A$ is determined by fitting the data points. (right) Distribution of the residuals of the data from the red curve in the left panel. The red curve is the best-fit Gaussian. 
} 
\end{figure}

Analyzing the the X-ray shadow image in Fig.~\ref{fig:mesh_image_det}, we examined the uniformity of the quantum efficiency. 
In Fig.~\ref{fig:mesh_qe} (left), we plot X-ray counts detected in regions corresponding to each of the $1.4~{\rm mm} \times 1.4~{\rm mm}$ holes of the mesh (Fig.~\ref{fig:mesh_photo}) against distance from the $^{55}{\rm Fe}$ source ($R$), which can be regarded as a point source. 
The data points follow well a $\propto R^{-2}$ relation. 
We show the distribution of the residuals from the $\propto R^{-2}$ curve in Fig.~\ref{fig:mesh_qe} (right). 
The distribution can be fit by a Gaussian with $\sigma = 0.038$. 
Considering the fact that the data points in Fig.~\ref{fig:mesh_qe} (left) typically have statistical errors of $\approx 3.0\%$, 
we conclude that the spatial variation of the quantum efficiency across the CCDs at the Mn-K$\alpha$ and K$\beta$ energies is at a $\sqrt{3.8^2 - 3.0^2} \sim 2\%$ level. 

\subsection{Charge Trail and CTI Corrections}
When charges are transferred in a CCD, a part of the charges may be left behind the pixels. 
We need to correct data for the effect called charge trail, otherwise the event is assigned a wrong grade, which 
leads to a misidentification of the event as a background. 
The charge trail correction is performed by using the method described by Nobukawa et al.\cite{nobukawa2014}
We note that different corrections should be applied between data taken with and without the window option 
because of different fast to slow vertical transfer number ratios between the two cases. 

\begin{figure}
\begin{center}
\begin{tabular}{c}
\includegraphics[width=8cm]{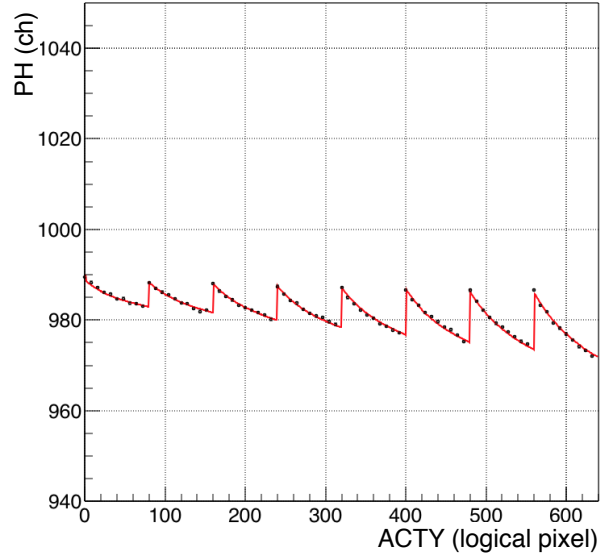}
\end{tabular}
\end{center}
\caption 
{\label{fig:cti_sawtooth_ccd3ab}PH of the Mn-K$\alpha$ line at 5.9~keV as a function of ACTY. 
The function shown with the red curve is the same one as used by Nobukawa et al.\cite{nobukawa2014}. 
The plotted data are from CCD3 Segment AB, where no CTI anomaly is found. 
} 
\end{figure}

\begin{figure}
\begin{center}
\begin{tabular}{c}
\includegraphics[width=16cm]{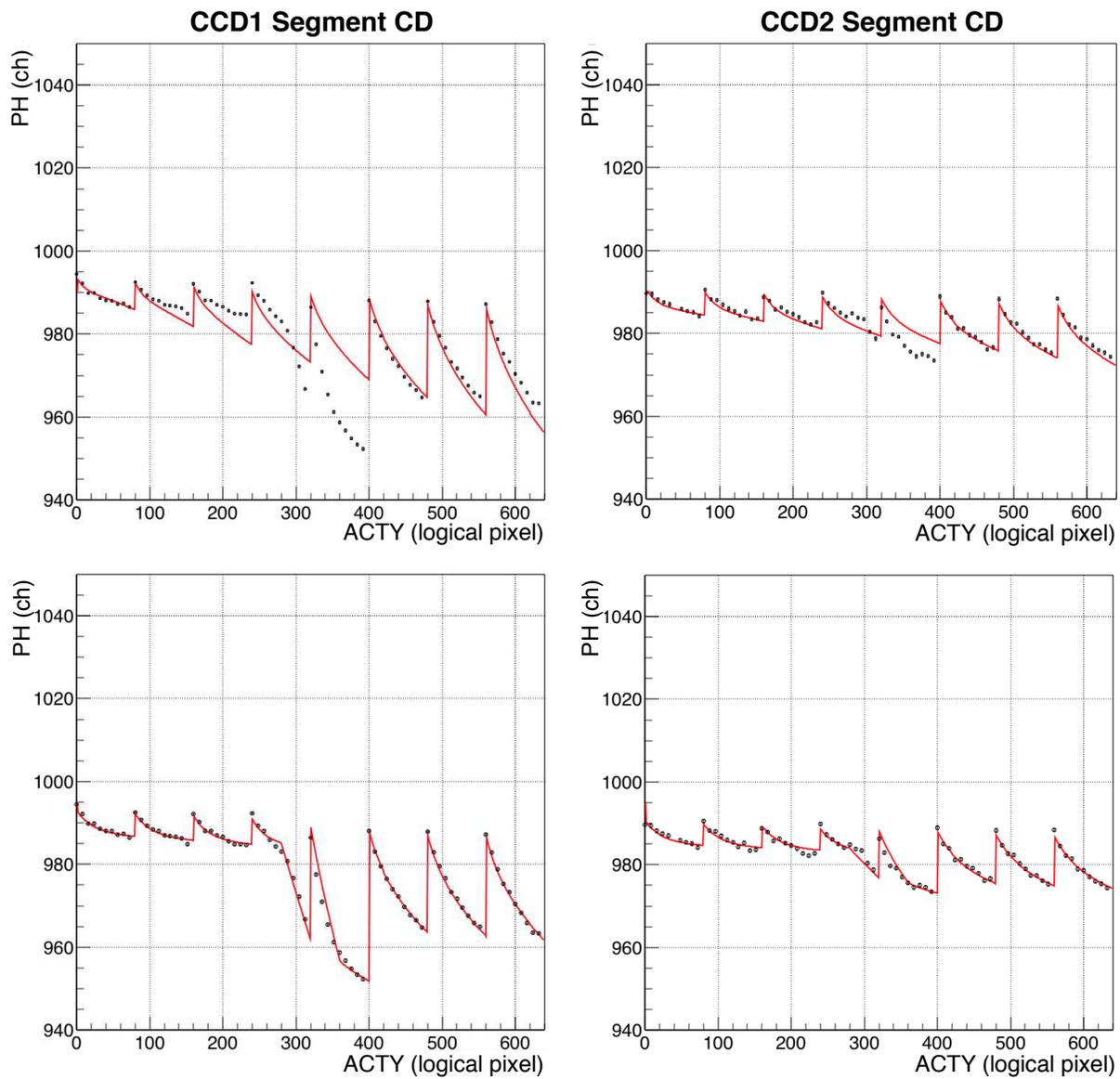}
\end{tabular}
\end{center}
\caption 
{\label{fig:cti_sawtooth}Same as Fig.~\ref{fig:cti_sawtooth_ccd3ab} but from CCD1 Segment CD (left panels) and CCD2 Segment CD (right panels). 
The analyzed regions are shown in Fig.~\ref{fig:dark_image}. The data are fit with the function proposed by Nobukawa et al.\cite{nobukawa2014} and the modified function 
(see text) in the top panels and bottom panels, respectively.  
} 
\end{figure}

\begin{figure}
\begin{center}
\begin{tabular}{c}
\includegraphics[width=12cm]{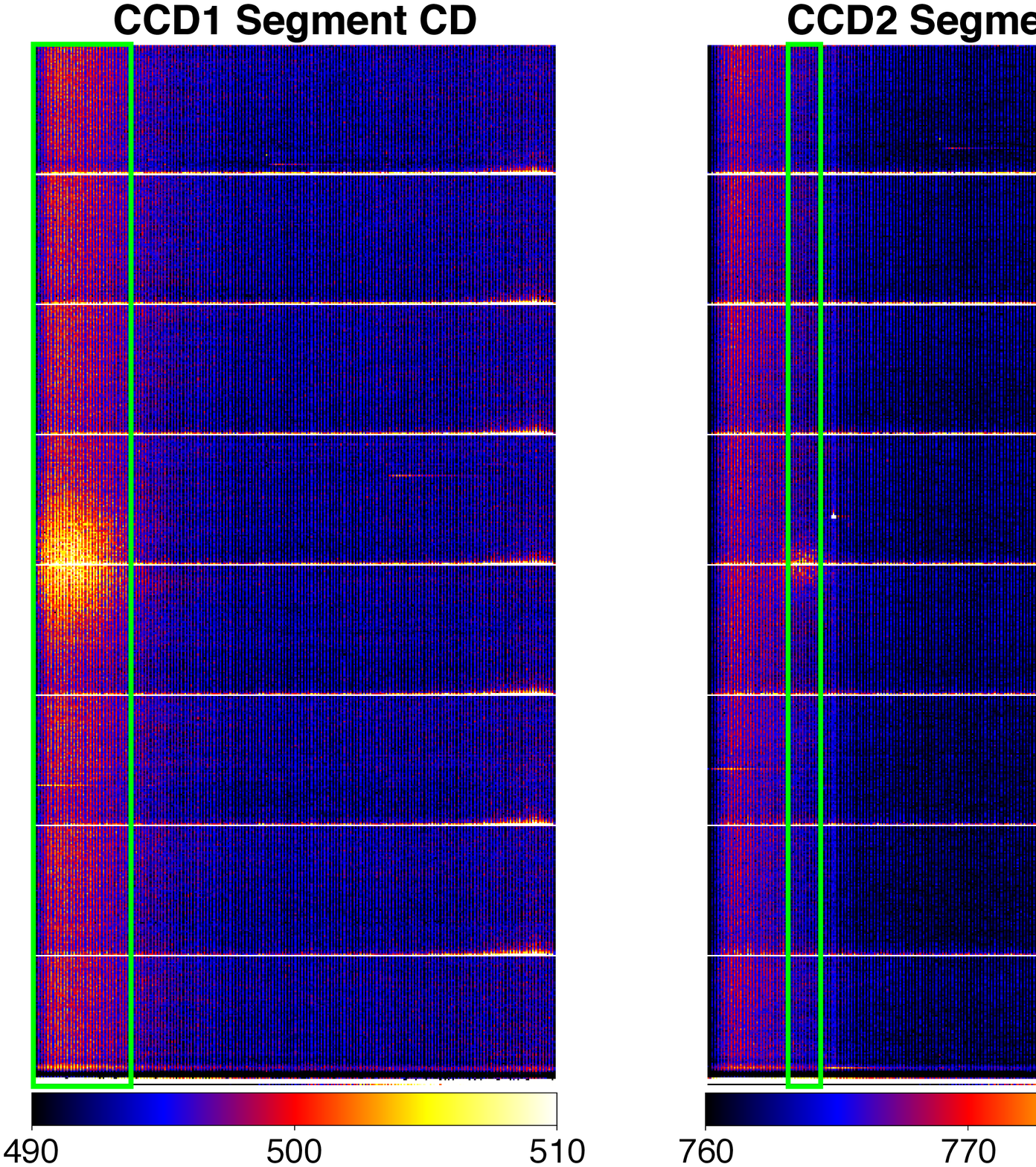}
\end{tabular}
\end{center}
\caption 
{\label{fig:dark_image} Dark frames from CCD1 Segment CD (left) and CCD2 Segment CD (right) taken at $-120~^\circ{\rm C}$. The green rectangles denote the regions used for the plots shown in Fig.~\ref{fig:cti_sawtooth}.  The rows in white are those with charge injection. 
} 
\end{figure}

We then apply CTI corrections.\footnote{According to our studies, the CTI during horizontal transfers are negligible. Thus, we focus on the CTI during vertical transfers in what follows.} 
In Fig.~\ref{fig:cti_sawtooth_ccd3ab}, we plot PH of 5.9~keV monochromatic  X-rays as a function of ACTY.
Because of the charge injection, a sawtooth shape appears in the plot. 
As demonstrated by Nobukawa et al.\cite{nobukawa2014},
the periodic shape can be fit by a function that considers injected charges filling traps and 
their re-emission, and PH can be corrected for the CTI. 
Data taken with the flight CCDs are generally well reproduced by the same function. 
However, we found that the function fails to fit data from some regions of the CCDs. 
We show examples in Fig.~\ref{fig:cti_sawtooth}. 
The CTI of some regions is significantly higher than the other locations of the CCDs. 
The locations of the anomaly coincide with those with higher dark level as shown in Fig.~\ref{fig:dark_image}. 
The method by Nobukawa et al.\cite{nobukawa2014} assumes all the parameters are uniform in a segment, 
and needs to be modified to reproduce the position-dependent CTI.

In order to take into account the CTI anomaly, we modified the correction method as follows. 
The relation between PH after the charge trail correction (${\rm PH}$) and CTI-corrected PH (${\rm PH}_0$) 
can be written as 
 \begin{eqnarray}\label{eq:cti}
{\rm PH} = {\rm PH}_0 \, (1-c_f)^{\rm Y_0} \, (1 - c_{f0})^{\rm Y_1} \, (1-c_a)^{\rm Y_2} \,  (1 - c_{a0})^{\rm Y_3} \, (1-c_s)^{\rm Y_4}.  
\end{eqnarray}
The parameter ${\rm Y_{\{0,1,2,3,4\}}}$ is a vertical transfer number with the definitions below. 
\begin{description}
\item[${\rm Y_0}$:] the number of fast transfers after passing the first charge injection row
\item[${\rm Y_1}$:] the number of fast transfers before reaching the first charge injection row 
\item[${\rm Y_2}$:] the number of fast transfers in CTI-anomaly regions after the first charge injection row 
\item[${\rm Y_3}$:] the number of fast transfers in CTI-anomaly regions before the first charge injection row 
\item[${\rm Y_4}$:] the number of slow transfers
\end{description}
We here assume that the CTI-anomaly regions are in the imaging area, not in the frame-store region, since data 
taken in on-ground tests are all consistent with the assumption. 
The parameters $c_f$, $c_a$, and $c_s$ refer to the CTI during fast transfers outside anomaly regions, fast transfers 
inside anomaly regions, and slow transfers. They are expressed as 
 \begin{eqnarray}\label{eq:cti_exp}
c_{\{f,a,s\}} = c_{\{f,a,s\}0} \left[1 - p_{\{f,a,s\}} \exp\left( -\frac{\Delta{\rm Y}}{\tau_{\{f,a,s\}}}\right) \right], 
\end{eqnarray}
where $c_{\{f,a,s\}0}$, $p_{\{f,a,s\}}$, $\tau_{\{f,a,s\}}$, and $\Delta{\rm Y}$ are the normalization, probability that 
a trap is filled by injected charges, de-trapping timescales of the charges in the unit of 
a transfer number, and difference of ACTY between the pixel considered and its preceding charge injection row. 
Please note that traps between the pixel and the charge injection row are not filled when the vertical transfer starts. 
That is why we have $c_{f0}$ and $c_{a0}$ instead of $c_{f}$ and $c_{a}$ in the third and fifth factors of Eq.~(\ref{eq:cti}).

\begin{figure}
\begin{center}
\begin{tabular}{c}
\includegraphics[width=8cm]{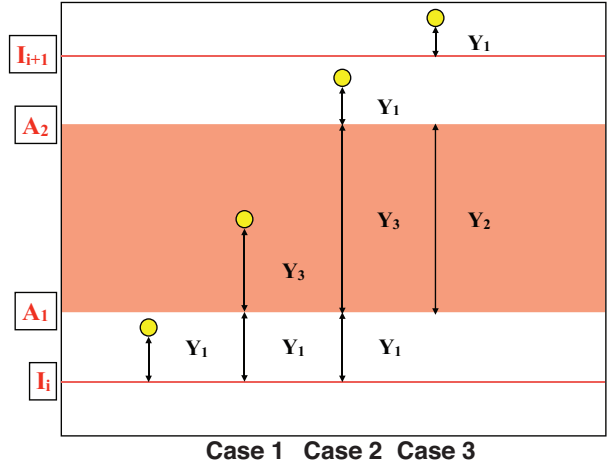}
\end{tabular}
\end{center}
\caption 
{\label{fig:cti_anomaly} Definitions of ${\rm Y_1}$, ${\rm Y_2}$, and ${\rm Y_3}$ in the case where a CTI anomaly (the shaded region) from ${\rm ACTY} = {\rm A}_1$ to ${\rm ACTY} = {\rm A}_2$ is between the two charge injection rows (the red horizontal lines) at ${\rm ACTY} = {\rm I}_{i}$ and ${\rm ACTY} = {\rm I}_{i+1}$
} 
\end{figure}

Let us explain in the following with an example shown in Fig.~\ref{fig:cti_anomaly}, in which a CTI anomaly region is located between two 
charge injection rows. 
The lower and upper boundaries of the CTI anomaly are at ${\rm ACTY} = {\rm A}_1$ and ${\rm ACTY} = {\rm A}_2$, respectively, and 
the charge injection rows are at ${\rm ACTY} =  {\rm I}_{0}, \cdots,  {\rm I}_{i},  {\rm I}_{i+1}, \cdots$.
If we operate the CCD with the full-window option and consider a pixel with ${\rm ACTY} = {\rm Y}$, 
the definition above gives ${\rm Y_0} = 640 -({\rm Y_1} + {\rm Y_2} + {\rm Y_3})$ and ${\rm Y_4} = {\rm Y}$. 
We can obtain ${\rm Y_1}$, ${\rm Y_2}$, and ${\rm Y_3}$ for each of the three cases below. 
\begin{description}
 \item[Case 1:] ${\rm A}_1  < {\rm Y} <  {\rm A}_2 $ 
 \begin{eqnarray}
 \begin{cases}
 {\rm Y}_1 =  {\rm A}_1 - {\rm I_i}  \\
 {\rm Y}_2 = 0 \\
 {\rm Y}_3 = {\rm Y} -  {\rm A}_1
  \end{cases}
 \end{eqnarray}
\item[Case 2:] ${\rm A}_2  < {\rm Y} <  {\rm I}_{i+1}$ 
 \begin{eqnarray}
 \begin{cases}
 {\rm Y}_1 = ({\rm Y}- {\rm A}_2) + ( {\rm A}_1 - {\rm I}_i) \\
 {\rm Y}_2 = 0 \\
 {\rm Y}_3 = {\rm A}_2 - {\rm A}_1
 \end{cases}
 \end{eqnarray}
\item[Case 3:] ${\rm I}_{i+1}  < {\rm Y}$ 
 \begin{eqnarray}
 \begin{cases}
 {\rm Y}_1 = \Delta{\rm Y} \\
 {\rm Y}_2 =  {\rm A}_2 - {\rm A}_1 \\
 {\rm Y}_3 = 0
 \end{cases}
 \end{eqnarray}
\end{description}
In a similar way, ${\rm Y_1}$, ${\rm Y_2}$, and ${\rm Y_3}$ can be calculated in other examples, in which, for example, 
a CTI anomaly region encompasses one or more charge injection rows. 

\begin{figure}
\begin{center}
\begin{tabular}{c}
\includegraphics[width=8cm]{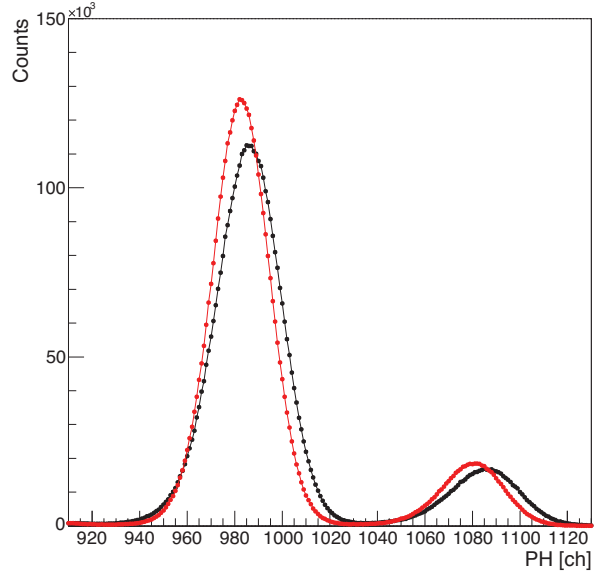}
\end{tabular}
\end{center}
\caption 
{\label{fig:spec_fe} Spectra of Mn-K$\alpha$ and K$\beta$ X-rays from $^{55}{\rm Fe}$ taken with CCD1 Segment CD before (black) and after (red) the trail and CTI corrections.  Events with grades 0, 2, 3, 4, and 6 are selected. 
Here the peaks are shifted toward lower energies after the corrections since ${\rm PH}_{0,\, 5.9~{\rm keV}}$ is set to $982~{\rm ch}$. 
} 
\end{figure}

The parameters in Eq.~(\ref{eq:cti_exp}) were determined by fitting data taken with the flight CCDs irradiated by 5.9~keV 
X-rays (Mn-K$\alpha$) from $^{55}{\rm Fe}$ in an on-ground test.  The fitting results are shown in Fig.~\ref{fig:cti_sawtooth}. 
The data points including those from the anomaly regions are well fit with the modified function. 
Energy dependence of CTI is taken into account by multiplying $c_{\{f,a,s\}}$ by a factor
\begin{eqnarray}
\left({\frac{{\rm PH}}{{\rm PH}_{0,\, 5.9~{\rm keV}}}}\right)^{-\beta_{\{f,a,s\}}}. 
\end{eqnarray}
Here ${\rm PH}_{0,\, 5.9~{\rm keV}}$ is ${\rm PH}_0$ at 5.9~keV, which is set to 982~ch. 
The index $\beta_{\{f,a,s\}}$ was determined using the F-K$\alpha$ line at 0.68~keV and Ge-K$\alpha$ line at 9.9~keV 
in addition to the 5.9~keV line. 
The indices are all in the range between $0.4$ and $0.8$. 
Figure~\ref{fig:spec_fe} shows $^{55}{\rm Fe}$ spectra from CCD1 Segment CD before and after the CTI correction. 
We applied the same grade selection as that used for Suzaku XIS, and extracted events with grades 0, 2, 3, 4, and 6. 
The energy resolution at 5.9~keV was improved from $194.2 \pm 0.8$~eV to $170.0 \pm 0.7$~eV in FWHM. 
The energy resolution was estimated by fitting the peak with a single Gaussian. 
Energy resolution of all the segments of the flight CCDs are summarized in Tab.~\ref{tab:fe_resolution}. 

\begin{table}[ht]
\caption{Energy resolution in FWHM for Mn-K$\alpha$ (5.9~keV) X-rays.} 
\label{tab:fe_resolution}
\begin{center}       
\begin{tabular}{ccc} 
\hline
\hline
\rule[-1ex]{0pt}{3.5ex}  CCD & Segment & Energy Resolution (eV)    \\
\hline 
\rule[-1ex]{0pt}{3.5ex}      CCD1 & AB & $167.6 \pm 0.7$ \\
\rule[-1ex]{0pt}{3.5ex}      CCD1 & CD & $170.0 \pm 0.7$ \\
\rule[-1ex]{0pt}{3.5ex}      CCD2 & AB & $165.5 \pm 0.7$ \\
\rule[-1ex]{0pt}{3.5ex}      CCD2 & CD & $163.6 \pm 0.7$ \\
\rule[-1ex]{0pt}{3.5ex}      CCD3 & AB & $164.2 \pm 0.7$ \\
\rule[-1ex]{0pt}{3.5ex}      CCD3 & CD & $165.0 \pm 0.7$ \\
\rule[-1ex]{0pt}{3.5ex}      CCD4 & AB & $160.7 \pm 0.7$ \\
\rule[-1ex]{0pt}{3.5ex}      CCD4 & CD & $161.2 \pm 0.7$ \\
\hline
\end{tabular}
\end{center}
\end{table}

\subsection{Detector Response}
\begin{figure}
\begin{center}
\begin{tabular}{c}
\includegraphics[width=8cm]{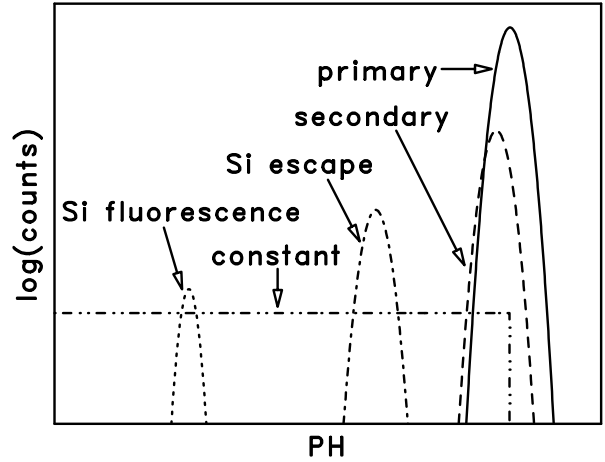}
\end{tabular}
\end{center}
\caption 
{\label{fig:response} Schematic picture of the response of the SXI CCDs.
} 
\end{figure}

Following the studies reported in literature\cite{bautz1999,XIS}, we modeled the CCD response as presented in Fig.~\ref{fig:response}. 
We decomposed the response to monochromatic X-rays with five components: a primary peak, 
a secondary peak, a constant tail, a Si-K fluorescence peak, and a Si escape peak. 
We can interpret them as follows. 
\begin{description}
\item[Primary peak] A Gaussian centered at the energy of the incident X-ray. 
This component corresponds to events in which an X-ray is absorbed and generated charges are fully detected. 
\item[Secondary peak] A Gaussian whose center energy is slightly lower than the primary peak. 
In this case, signal charges are split into more than one pixel, and a part of the charges are not detected 
 since the PH is lower than the split threshold. 
\item[Constant] When an X-ray is absorbed close to the interface between the depletion layer and the oxide layer 
and only a part of the charges are collected, the event constitutes this component. 
\item[Si-K fluorescence peak] 
The peak is attributed to events in which an X-ray is absorbed in the depletion layer away from the corresponding pixel 
or in an inactive part of the CCD such as the oxide layer, and then the escaped Si fluorescence X-ray photon is 
detected in the pixel. 
\item[Si escape peak] 
This is the case in which an X-ray is absorbed and then the Si fluorescence X-ray escapes from the pixel. 
\end{description}

\begin{figure}
\begin{center}
\begin{tabular}{c}
\includegraphics[width=16cm]{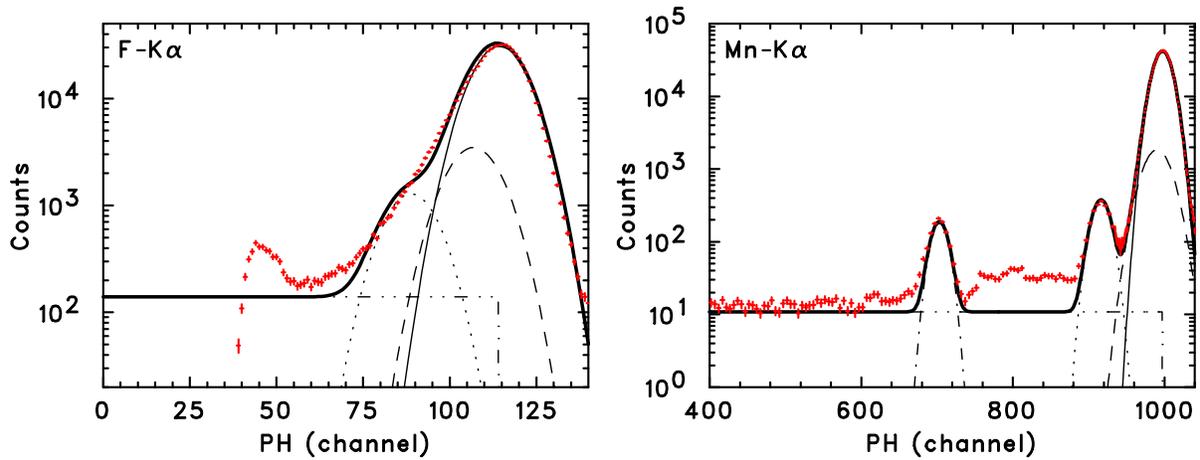}
\end{tabular}
\end{center}
\caption 
{\label{fig:response_fit} 
Spectra of the F-K$\alpha$ line at 0.68~keV (left) and the Mn-K$\alpha$ line at 5.9~keV (right) taken with CCD4 Segment CD in an on-ground test. 
There are no counts below $\approx 40$~channel since the threshold for event detection was set to 40 channel. 
The thin solid, dashed, dash-dot-dot, dash-dot curves represent the primary peak, secondary peak, constant, Si escape peak components, respectively. 
The dotted curves are Gaussians to account for background lines: the O-K$\alpha$ in the left panel and the Cr-K$\alpha$ line in the right panel. 
The thick solid curves indicate the sum of all the components. 
Significant residuals can be explained by minor background lines which are not included in the model here. 
The excess peaking at $\sim 40$~ch in the left panel are due to the C-K$\alpha$ line. 
The excess from $\sim 750$~ch to $\sim 880$~ch in the right panel can be explained by a blend of the Ti-K$\alpha$ line and the Si escape peak 
associated with the Mn-K$\beta$ line. 
} 
\end{figure}

\begin{figure}
\begin{center}
\begin{tabular}{c}
\includegraphics[width=16cm]{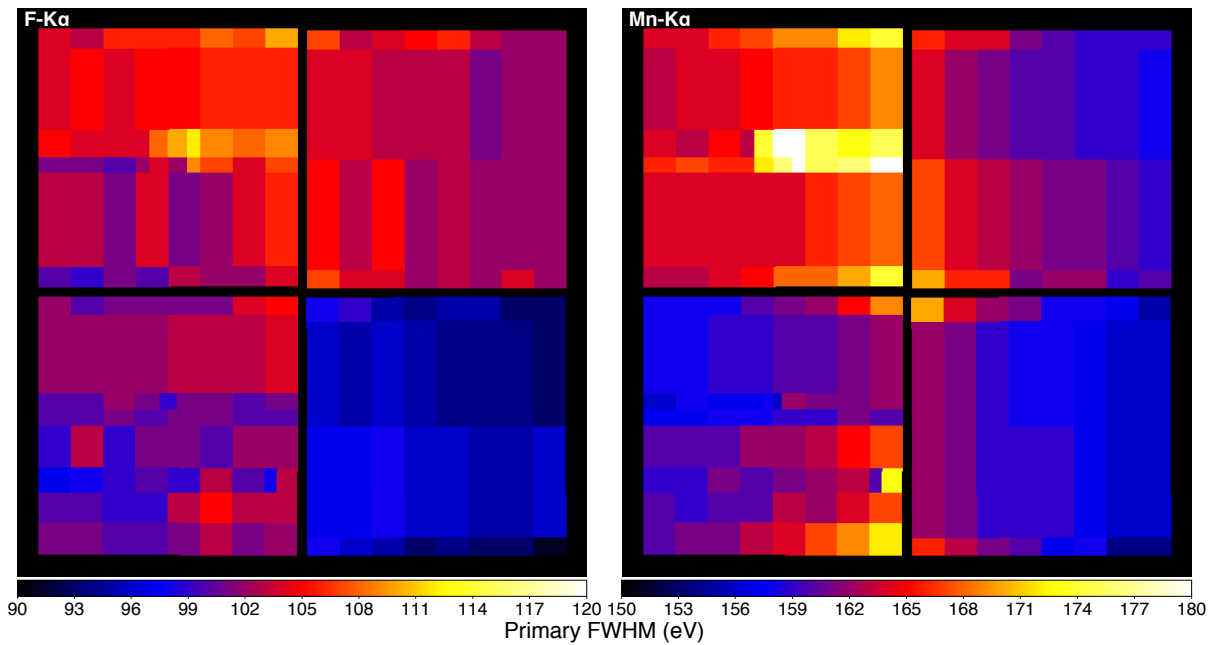}
\end{tabular}
\end{center}
\caption 
{\label{fig:primary_fwhm} Maps of widths in FWHM of the primary peak at 0.68~keV (left) and at 5.9~keV (right) in the DET coordinate. This result is reflected in the detector response function used for observational data analysis. 
} 
\end{figure}

In Fig.~\ref{fig:response_fit}, we show spectra taken in an on-ground test overlaid with the model curves convolved with the response function. 
We determined parameters for each of the above components based on data taken in on-ground tests as well as Monte Carlo simulations. 
We modeled the constant to primary peak intensity ratio according to Monte Carlo simulations as already reported elsewhere\cite{inoue2016}. 
The secondary peak to primary peak intensity ratio is phenomenologically modeled to fit on-ground test results. 
The ratio decreases with energy: 0.12 at 0.68~keV and 0.066 at 5.9~keV. 
We took into account the spatial variation of the energy resolution 
in each CCD. This is particularly important to reproduce spectra from the CTI-anomaly regions 
since energy resolution there is also worse than that of the other regions.  
We evaluated FWHM of the primary peak in each region of the CCDs using F-K$\alpha$, Ge-L$\alpha$, Mn-K$\alpha$, 
and Ge-K$\alpha$ lines. 
The energy resolution maps at the F-K$\alpha$ and Mn-K$\alpha$ line energies are presented in Fig.~\ref{fig:primary_fwhm}.

\subsection{Optical and Ultraviolet Light Blocking}
X-ray CCDs are sensitive also to optical and ultraviolet (UV) photons, and thus the SXI has OBL and CBF to block them. 
The SXI is required to perform an observation even with a star as bright as an apparent magnitude of $1.0$ in the FoV. 
This leads to an optical light transmission of $< 10^{-5}$ as a requirement. 
In the UV band, the most stringent requirement comes from the HeII line emission at $30.4~{\rm nm} \, (= 40.8~{\rm eV})$ 
through resonance scattering of sunlight in the geocorona. 
The intensity of the emission, $\sim 10^{7}~{\rm photons}~{\rm cm}^{-2}~{\rm s}^{-1}~{\rm sr}^{-1}$, demands a transmission 
of $< 10^{-2}$ at the wavelength. 

\begin{figure}
\begin{center}
\begin{tabular}{c}
\includegraphics[width=12cm]{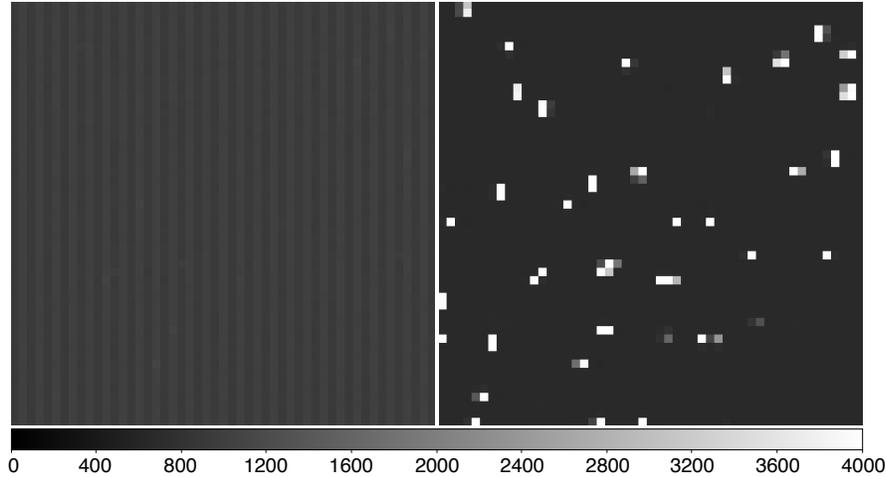}
\end{tabular}
\end{center}
\caption 
{\label{fig:dark_comparison} Dark frames taken in May 2013 (left) and September 2014 (right) with optical light illumination . Shown are the same $\sim 50 \times 50$ logical pixel region 
of CCD4 Segment AB. The white pixels in the right panel are those with pinholes in the OBL. The pixel values reflect the optical light transmission in linear scale. 
} 
\end{figure}

We measured the transmission of optical light through the OBL using a prototype CCD, 
confirming that the OBL satisfied the requirement for the optical transmission of $< 10^{-5}$. 
During on-ground tests of the flight hardware, however, we found a number of pinholes in the OBL developing with time. 
Fig.~\ref{fig:dark_comparison} shows dark frames from one of the four CCDs illuminated by optical light from an LED, taken about a year apart. 
At the time of the latter measurement, 3--5\% of the pixels have pinholes, and the optical transmission was estimated to be 
$\sim 100$--$1000$ times higher at those pixels. 
We, therefore, changed the specifications of the CBF so that optical light can be blocked by the CBF instead of the OBL. 
We vapor-deposited Al on both sides of the CBF with thicknesses of 80~nm and 40~nm in the new design 
whereas only one side was coated with 30-nm thick Al in the original design. 
We confirmed that the optical transmission of the CBF is  $< 10^{-5}$ through measurements of samples equivalent to the flight model. 
UV photons at 40.8~eV are mainly absorbed by 200-nm thick polyimide layer of the CBF. 
We measured UV transmission of the CBF in the energy range between $38~{\rm eV}$ and $60~{\rm eV}$ using 
the same samples as those used for the optical transmission measurement. 
The transmission obtained is  $\sim 3 \times 10^{-4}$, $\sim 6 \times 10^{-4}$, and $\sim 2 \times 10^{-2}$ at $38~{\rm eV}$, $40~{\rm eV}$, 
and $60~{\rm eV}$, respectively, which satisfies the above mentioned requirement.

\section{Summary}
Hitomi SXI is an imaging spectrometer, covering the energy range between $0.4~{\rm keV}$ and $12~{\rm keV}$. 
The sensors of the SXI are fully-depleted, back-illuminated CCDs with a $200~\mu{\rm m}$-thick depletion layer. 
We arrange the CCDs in a $2 \times 2$ array. The total imaging area size is $\sim 60~{\rm mm} \times 60~{\rm mm}$, which 
corresponds to a FoV of $38^{\prime} \times 38^{\prime}$. 
The CCDs can be cooled by Stirling coolers down to $-120~^\circ{\rm C}$ so that we can mitigate CTI due to in-orbit radiation damage. 
To achieve even lower CTI, we adopt a charge injection technique similar to that used for Suzaku XIS. 
SXI-PE and SXI-DE perform on-board data processing such as dark subtraction and event extraction.  
SXI-PE also generates CCD clock patterns which can be programmed flexibly by using microcodes. 
On-ground tests confirm the imaging and spectroscopic capability of the SXI. 
The energy resolution is measured to be $161$--$170~{\rm eV}$ in FWHM for 5.9~keV X-rays. 
During the tests, we found two problems. One is that CTI of some regions of the CCDs are found to be significantly higher. 
We developed a method to properly correct for the position-dependent CTI. 
The other problem is the pinholes in the OBL. 
We increased the Al thickness of the CBF so that optical light can be blocked by the CBF instead of the OBL. 

\acknowledgments 
The authors thank all the former team members including former graduate students for their contribution to the development of the SXI.  
This work was supported by JSPS/MEXT KAKENHI Grant Numbers JP24684010, JP26670560, JP15H03641, JP15K17610, 
JP16H03983, JP23000004, JP16K13787, JP16H00949, JP26109506, JP23340071, JP21659292, JP26800144, JP24740123, 
JP25870347, JP25109004, JP26800102, JP15H02090, JP14079204, JP20365505, JP23740199, JP18740110, JP20549005, 
JP25870181, JP16J00548, JP17K14289, and JP15J01845. 
SI and KKN are supported by Research Fellowships of the JSPS for Young Scientists. 
X-ray transmission of the CBF was measured at  BL25-SU of SPring-8 
with the approval of the Japan Synchrotron Radiation Research Institute (JASRI) (Proposal Number 2013B1194). 
The measurements of X-ray and EUV transmission of the CBF and OBL as well as the measurements of quantum efficiency 
and soft X-ray response of the CCD 
were performed at BL-11A, 11B, and 20A of the Photon Factory at the High Energy Accelerator Research Organization (KEK) 
(Proposal Numbers 2010G594, 2010G686, 2012G621, 2012G622, 2013G593, 2014G607, 2015G627, and 2016G655).



\bibliography{sxi_reference}  
\bibliographystyle{spiejour}   


%
%

\end{spacing}
\end{document}